\documentclass[aps,prb,twocolumn,superscriptaddress,floatfix,letterpaper]{revtex4}

\pdfoutput=1

\usepackage[bbgreekl]{mathbbol}
\usepackage{amscd}

\newcommand{\calC}{\mathcal{C}}

\newcommand{\Z}{\mathbb{Z}}

\usepackage{tikz}

\begin{document}

\title{
A classification of 2D fermionic and bosonic topological order
}

\author{Zheng-Cheng Gu}
\affiliation{Perimeter Institute for Theoretical Physics, Waterloo, Ontario, N2L 2Y5 Canada} 

\author{Zhenghan Wang}
\affiliation{Microsoft Station Q, CNSI Bldg. Rm 2237, University of California, Santa Barbara, CA 93106}

\author{Xiao-Gang Wen}
\affiliation{Department of Physics, Massachusetts Institute of Technology, Cambridge, Massachusetts 02139, USA}
\affiliation{Perimeter Institute for Theoretical Physics, Waterloo, Ontario, N2L 2Y5 Canada} 

\date{May, 2010}
\begin{abstract}
The string-net approach by Levin and Wen, and the local unitary transformation
approach by Chen, Gu, and Wen, provide ways to classify topological orders with
gappable edge in 2D bosonic systems. The two approaches reveal that the
mathematical framework for 2+1D bosonic topological order with gappable edge is
closely related to unitary fusion category theory.  In this paper, we
generalize these systematic descriptions of topological orders to 2D fermion
systems.  We find a classification of 2+1D fermionic topological orders with
gappable edge in terms of the following set of data $(N^{ij}_k, F^{ij}_k,
F^{ijm,\al\bt}_{jkn,\chi\del},d_i)$,
that satisfy a set of non-linear algebraic equations.  The exactly soluble
Hamiltonians can be constructed from the above data on any lattices to realize
the corresponding topological orders.  When $F^{ij}_k=0$, our result recovers
the previous classification of 2+1D bosonic topological orders with gappable
edge.

\end{abstract}

\maketitle

{\small \setcounter{tocdepth}{1} \tableofcontents }

\section{Introduction}

Understanding phases of matter is one of the central problems
in condensed matter physics.  Landau symmetry breaking
theory,\cite{L3726,LanL58} as a systematic theory of phases
and phase transitions, becomes a cornerstone of condensed
matter theory. However, at zero temperature, the symmetry
breaking states described by local order parameters are
basically direct product states. It is hard to believe that
various direct product states can describe all possible
quantum phases of matter.

Based on adiabatic evolution, one can show that gapped quantum
phases at zero temperature correspond to the equivalence
classes of local unitary (LU) transformations generated by
finite-time evolutions of local hermitian operators $\t
H(\tau)$:\cite{HW0541,BHV0601,BHM1044,CGW1038,ZW1490}
\begin{align}
 |\Phi\>\sim |\Phi'\>
\ \ \ \text{ iff } |\Phi'\>=\cT \e^{\imth \int d\tau \t H(\tau)} |\Phi\>
.
\end{align}
It turns out that there are many gapped quantum states that cannot
be transformed into direct product states through LU
transformations.  Those states are said to have a long range
entanglement.  Thus, the equivalence classes of LU
transformations, and hence the quantum phases of matter, are
much richer than direct product states and  much richer than
what the symmetry breaking theory can describe.  Different
patterns of long range entanglement correspond to different
quantum phases that are beyond the
symmetry-breaking/order-parameter description\cite{W0275}
and direct-product-state description.  The patterns of long
range entanglement really correspond to the topological
orders\cite{Wrig,Wtoprev} that describe the new kind of
orders in quantum spin liquids and quantum Hall
states.\cite{RK8876,WWZcsp,RS9173,Wsrvb,MS0181,Wqoslpub,K032,MR9162,Wnab}

In absence of translation symmetry, the above LU
transformation can be expressed as a quantum circuit, which
corresponds to a discretized LU transformation.  The
discretized LU transformation is more convenient to use.
The gapped quantum phases can be more effectively studied and
even be classified through the discretized LU
transformations.\cite{LWstrnet,VCL0501,V0705,CGW1038,ZW1490}

After discovering more and more kinds of topological orders,
it becomes important to gain a deeper understanding of
topological order under a certain mathematical framework.
We know that symmetry breaking orders can be understood
systematically under the mathematical framework of group
theory.  Can topological orders be also understood under
some mathematical framework? From the systematic
construction of topologically ordered states based on
string-nets\cite{LWstrnet} and the systematic description of
non-Abelian satistics\cite{K062}, it appears that tensor
category theory may provide the underlying mathematical
framework for topological orders.\cite{FNS0428}

However, the string-net and the LU transformation
approaches\cite{LWstrnet,CGW1038,ZW1490} only provide a systematic
understanding for topological orders in qubit systems (\ie
quantum spin systems or local boson systems). Fermion
systems can also have non trivial topological orders. In
this paper, we will introduce a systematic theory for
topological orders in interacting fermion systems (with
interacting boson systems as special cases).  Our approach
is based on the LU transformations generated by local
hermitian operators that contain even number of fermion
operators. It allows us to classify and construct a large
class of topological orders in fermion systems. The
mathematical framework developed here may be related to the
theory of enriched categories,\cite{K051} which can be
viewed as a generalization of the standard tensor category
theory\cite{Mac71,Wang10}

To gain a systematic understanding of topological order in
fermion systems, we first need a way to label those
fermionic topological orders. In this paper, we show that a
large class of fermionic topological orders (which include
bosonic topological orders as special cases) can be labeled
by a set of tensors: $(N^{ij}_k, F^{ij}_k,
F^{ijm,\al\bt}_{jkn,\chi\del},d_i)$.
Certainly, not every set of tensors corresponds to a valid
fermionic topological order.  We show that only the tensors
that satisfy a set of non-linear equations
correspond to valid fermionic topological orders.  The set
of non-linear equations obtained here is a
generalization of the non-linear equations (such as the
pentagon identity) in a tensor category theory.  So our
approach is a generalization of tensor category theory and
the string-net approach for bosonic topological orders.  We would
like to point out that the framework developed here not
only leads to a classification of fermionic topological
orders, it also leads to a more general classification of
bosonic topological orders than the string-net and the
related approaches.\cite{LWstrnet,CGW1038}

{}From a set of the data $(N^{ij}_k, F^{ij}_k,
F^{ijm,\al\bt}_{jkn,\chi\del},d_i)$,
we can obtain the parent Hamiltonian as a
sum of projectors. We believe that the Hamiltonian is
unfrustrated.  Its zero-energy ground state realizes the
fermionic/bosonic topological order described by the data.

In section II, we give a careful discussion on what is a
local fermion system.  In section III, we introduce
fermionic local unitary transformations. We then use
fermionic local unitary transformations to define quantum
phases for local fermion systems and fermionic topological
orders. In section IV, we use fermionic local unitary
transformations to define a wave function renormalization
flow for local fermion systems. In section V, we discuss
the fixed points of the wave function renormalization flow,
and use those fixed points to classify a large class of
fermionic (and bosonic) topological orders.  
In section VI, we comment on its relation to categorical
framework.  In section VII, we give a few simple examples.
In appendix A, we discuss the definition of branching structure for a trivalent graph. 
In appendix B, we discuss the fermionic structure of the
support space.  In appendix C, we derive the ideal
Hamiltonian from the data that characterize the
fermionic/bosonic topological orders.  

\section{Local fermion systems}

Local boson systems (\ie local qubit systems) and local
fermion systems have some fundamental differences. To reveal
those differences, in this section, we are going to define
local fermion systems carefully.  To contrast local fermion
systems with local boson systems, let us first review the
definition of local boson systems.

\subsection{Local bosonic operators and bosonic states in a
local boson model}

A local boson quantum model is defined through its Hilbert
space $V$ and its local boson Hamiltonian $H$.  The Hilbert
space $V$ of a local boson quantum model has a structure
$V=\otimes V_{\v i}$ where $V_{\v i}$ is the local Hilbert space
on the site $\v i$.  A local bosonic operator is
defined as an operator that act within the local Hilbert
space $V_{\v i}$, or as a finite product of local bosonic
operators acting on nearby sites.  A local boson Hamiltonian
$H$ is a sum of local bosonic operators.
The ground state of a local boson Hamiltonian $H$
is called a bosonic state.

\subsection{Local fermionic operators and fermionic states
in a local fermion model}

Now, let us try to define local fermion systems.
A \emph{local fermion quantum model} is also defined through its
Hilbert space $V$ and its local fermion Hamiltonian $H_f$.
First let us introduce \emph{fermion operator} $c^\al_{\v i}$ at
site $\v i$ as operators that satisfy the
anticommutation relation
\begin{align}
\label{antic}
 c^\al_{\v i} c^\bt_{\v j} = - c^\bt_{\v j} c^\al_{\v i},\ \ \ \ \
 c^\al_{\v i} (c^\bt_{\v j})^\dag = - (c^\bt_{\v j})^\dag c^\al_{\v i},
\end{align}
for all $\v i\neq \v j$ and all values of the $\al,\bt$ indices.
We also say $c^\al_{\v i}$ acts on the site $\v i$.
The fermion Hilbert space $V$ is the space generated by
the fermion operators and their hermitian
conjugate:
\begin{align}
 V=\{[c^\bt_{\v j}(c^\al_{\v i})^\dag ...] |0\>\} .
\end{align}
Due to the anticommutation relation \eq{antic}, $V$ has a
form $V=\otimes V_{\v i}$ where $V_{\v i}$ is the local Hilbert
space on the site $\v i$. We see that the total Hilbert
space of a fermion system has the same structure as a
local boson system.

Using the Hilbert space  $V=\otimes V_{\v i}$, an explicit
representation of the fermion operator $c^\al_{\v i}$ can be
obtained.  First, each local Hilbert space can be splitted as
$V_{\v i}=V^0_{\v i}\oplus V_{\v i}^1$.
We also choose an ordering of the site label $\v i$.
Then $c^\al_{\v i}$ has the following matrix representation
\begin{align}
 c^\al_{\v i} &=
C^\al_{\v i} \prod_{\v j<\v i} \Si^3_{\v j},
\nonumber\\
C^\al_{\v i}&=\bpm
0 & A^\al_{\v i}\\
B^\al_{\v i} & 0\\
\epm, \ \ \ \
\Si^3_{\v i}=\bpm
I^0_{\v i} & 0\\
0 & -I^1_{\v i}\\
\epm ,
\end{align}
where $I^0_{\v i}$ is the identity matrix acting in the space
$V^0_{\v i}$ and $I^1_{\v i}$ is the identity matrix acting in the
space $V^1_{\v i}$.  The matrix $C^\al_{\v i}$ maps a state in
$V^0_{\v i}$ to a state in $V^1_{\v i}$, and vice versa.
We note that
\begin{align}
C^\al_{\v i} \Si^3_{\v i}  = -\Si^3_{\v i}  C^\al_{\v i} .
\end{align}
We see that a fermion operator is not a local bosonic operator.  The product of
an odd number of fermion operators and any number of local bosonic operators on
nearby sites is called a \emph{local fermionic operator}.

Let us write the eigenvalue of $\Si^3_{\v i}$ as
$(-)^{s_{\v i}}$.  The states in $V^0_{\v i}$ have $s_{\v
i}=0$ and are called bosonic states.  The states in
$V^1_{\v i}$ have $s_{\v i}=1$ and are called fermionic
states. We can view $s_{\v i}$ as the fermion number
on site $\v i$.

A \emph{local fermion Hamiltonian} $H_f$ is a sum of terms:
$H=\sum_P O_{P}$, where $\sum_P$ sums over a set of regions.
Each term $O_P$ is a product of an even number of local
fermionic operators and any number of local bosonic operators
on a finite region $P$.  Such kind of terms is called
\emph{pseudo-local bosonic operator} acting on the region.
In other words, a local fermion Hamiltonian is a sum of
pseudo-local bosonic operators.  The ground state of a local
fermion Hamiltonian $H_f$ is called a \emph{fermionic
state}.

Note that, beyond 1D, a pseudo-local bosonic operator is in
general not a local bosonic operator.  So a local fermion
Hamiltonian $H_f$ (beyond 1D) in general is not a local
boson Hamiltonian defined in the last subsection.  In this
sense, a local boson system and a local fermion system are
fundamentally different despite they have the same Hilbert
space. When viewed as a boson system, a local fermion
Hamiltonian corresponds to a non-local boson Hamiltonian
(beyond 1D).  Thus classifying the quantum phases of local
fermion systems corresponds to classifying the quantum
phases of a particular kind of non-local boson systems.

\section{
Fermionic local unitary transformation
and topological phases of fermion systems
}

Similar to the local boson systems, the finite-time evolution
generated by a local fermion Hamiltonian defines an
equivalence relation between gapped fermionic states:
\begin{align}
\label{fLUdef}
 |\psi(1)\> \sim |\psi(0)\> \text{\ iff\ }
 |\psi(1)\> =  \cT[\e^{\imth \int_0^1 dg\, \t H_f(g)}] |\psi(0)\>
\end{align}
where $\cT$ is the path-ordering operator and $\t
H(g)=\sum_{\v i} O_{\v i}(g)$ is a local fermion Hamiltonian
(\ie $O_{\v i}(g)$ is a pseudo-local bosonic operator which
is a product of even local fermionic operators).  We will
call $ \cT[\e^{\imth \int_0^1 dg\, \t H_f(g)}] $ a fermion local
unitary (fLU) evolution.  We believe that the equivalence
classes of such an equivalence relation are the universality
classes of the gapped quantum phases of fermion systems.

The finite-time fLU evolution introduced here is closely
related to \emph{fermion quantum circuits with finite
depth}.  To define fermion quantum circuits, let us
introduce  piecewise fermion local unitary operators.  A
piecewise fermion local unitary operator has a form
$ U_{pwl}= \prod_{i} \e^{\imth  H_f(\v i)}$,
where $\{ H_f(\v i) \}$ is a Hermitian operator which is a
pseudo-local bosonic operator that acts on a region labeled by $\v
i$. Note that regions labeled by different $\v i$'s are not
overlapping.  $U_{\v i} =\e^{\imth  H_f(\v i)}$ is called a
fermion unitary operator.  The size of each region is less
than some finite number $l$. The unitary operator $U_{pwl}$
defined in this way is called a fermion piece-wise local
unitary operator with range $l$.  A fermion quantum
circuit with depth $M$ is given by the product of $M$
fermion piece-wise local unitary operators:
$U^M_{circ}= U_{pwl}^{(1)} U_{pwl}^{(2)} \cdots
U_{pwl}^{(M)}$.
We believe that finite time fLU evolution can be
simulated with a constant depth fermion quantum circuit and
vice versa.  Therefore, the equivalence relation
\eqn{fLUdef} can be equivalently stated in terms of constant
depth fermion quantum circuits:
\begin{equation}
\label{PhiUcPhi}
|\psi(1)\> \sim |\psi(0)\> \text{ iff }
 |\psi(1)\> = U^M_{circ} |\psi(0)\>
\end{equation}
where $M$ is a constant independent of system size. Because
of their equivalence, we will use the term ``fermion Local
Unitary Transformation'' to refer to both fermion local
unitary evolution and constant depth fermion quantum
circuit in general.

Just like boson systems, the equivalence classes of
fermionic local unitary transformations correspond to the
universality classes that define phases of matter.  Since
here we do not include any symmetry, the equivalence classes
actually correspond to topologically ordered phases.  Such
topologically ordered phases will be called fermionic
topologically ordered phases.

\section{
Fermionic local unitary transformation
and wave function renormalization
}

After defining the fermionic topological orders as the
equivalence classes of many-body wave functions under fLU
transformations, we like to use the fLU transformations, or
more precisely the generalized fermion local unitary (gfLU)
transformation, to define a wave function renormalization
procedure. The wave function renormalization can remove the
non-universal short-range entanglement and make generic
complicated wave functions to flow to some very simple
fixed-point wave functions.  The simple fixed-point wave
functions can help us to classify fermionic topological
orders.

\begin{figure}
\begin{center}
\includegraphics[scale=0.45]{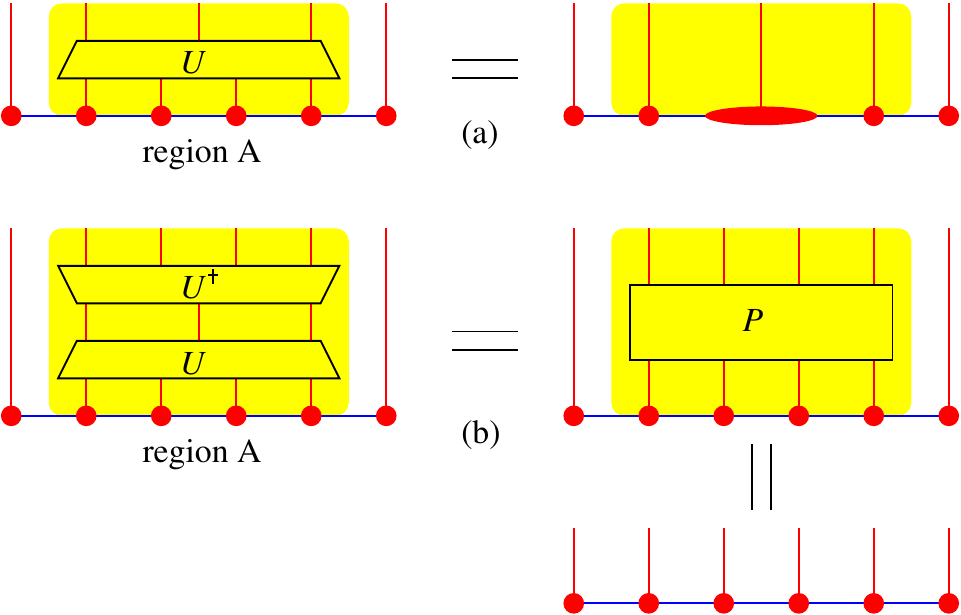}
\end{center}
\caption{
(Color online)
(a) A fermion local unitary (gfLU) transformation $U_g$
acts in region A of a fermionic state $|\psi\>$ are formed
by bosonic and fermionic operators in the region A. $U_g$
always contain even numbers of fermionic operators.  (b)
$U_g^\dag U_g=P$ is a projector, whose action does not
change the state $|\psi\>$.
}
\label{gLUT}
\end{figure}

Let us first define the gfLU transformation $U_g$ more
carefully.
Consider a state $|\psi\>$.  Let $\rho_A$ be the
entanglement density matrix of $|\psi\>$ in region $A$.
$\rho_A$ may act in a subspace of the Hilbert space in
region A. The subspace is called the support space $\t V_A$
of region $A$ (see Fig.  \ref{gLUT}(a)).  Let $|\t \phi_{\t
i}\>$ be a basis of this support space $\t V_A$, and
$|\phi_{i}\>$ be a basis of the full Hilbert space $V_A$ of
region $A$.  The gfLU transformation $U_g$ is the projection from the
full Hilbert space $V_A$ to the support space $\t V_A$.
So up to some unitary transformations, $U_g$ is a hermitian
projection operator:
\begin{align}
 \label{UUp}
U_g &=U_1P_gU_2, & P_g^2&=P_g, & P_g^\dag&=P_g,
\nonumber\\
U_1^\dag U_1&=1, & U_2^\dag U_2 &=1.
\end{align}
The matrix elements of $U_g$ are given by $\<\t \phi_{\t
a}|\phi_i\>$.  We will call such a gfLU transformation a
primitive gfLU transformation. A generic  gfLU
transformation is a product of several  primitive gfLU
transformations which may contain several hermitian
projectors and unitary transformations, for example,
$U_g =U_1P_gU_2P'_gU_3$.

To understand the fermionic structure of $U_g$, we note that
the support space $\t V_A$ has a structure $ \t V_A = \t
V_A^0 \oplus \t V_A^1 $ (see appendix \ref{fspsp}), where
$\t V_A^0$ has even numbers of fermions and $\t V_A^1$ has
odd numbers of fermions.  This means that $U_g$ contains
only even numbers of fermionic operators (\ie $U_g$ is a
pseudo-local bosonic operator).

We also regard the inverse of $U_g$, $U_g^\dag$, as a gfLU
transformation.  An fLU transformation is viewed as a special
case of gfLU transformations where the degrees of freedom
are not changed.  Clearly $U_g^\dag U_g=P$ and $U_g
U_g^\dag=P'$ are two hermitian projectors.  The action of $P$ does not
change the state $|\psi\>$ (see Fig. \ref{gLUT}(b)).  Thus
despite the degrees of freedom can be reduced under the gfLU
transformations, no quantum information of the state $|\psi\>$ is
lost under the gfLU transformations.

We note that the gfLU transformations
can map one wave function to another wave function with fewer
degrees of freedom. Thus it can be viewed as a wave function
renormalization group flow.  If the wave function
renormalization leads to fixed-point wave functions, then
those fixed-point wave functions can be much simpler, which
can provide an efficient or even one-to-one labeling
scheme of fermionic topological orders.

\section{
Wave function renormalization
and a classification of fermionic topological orders
}
\label{fixwv}

As an application of the above fermionic wave function renormalization, in
this section, we will study the structure of fixed-point
wave functions under the wave function renormalization.
This will lead to a classification of fermionic topological orders.

\subsection{Quantum state on a graph}

\begin{figure}[tb]
\begin{center}
\includegraphics[scale=0.6]{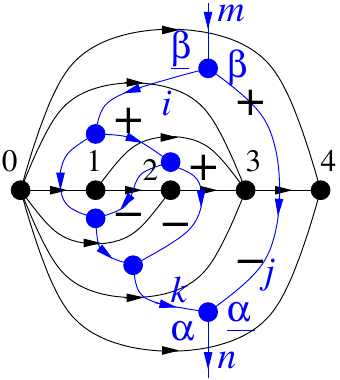}
\end{center}
\caption{
The lattice is a graph (described by the blue lines) with branching structure.
The black lines describe the dual lattice.  The state on the edges are labeled
by $i,j,k,m,n=0,\cdots,N$.  The state on the edges are labeled by $\al,\bt$.
The vertices are labeled by $\u\al,\u\bt$.  The $\u\al$ vertex has two incoming
edges and $\al$ has a range $\al=1,\cdots,N^{kj}_n$.  The $\u\bt$ vertex has
two incoming edges and $\bt$ has a range $\bt=1,\cdots,N_{ij}^m$.
}
\label{tri2DC}
\end{figure}

Since the wave function renormalization may change the lattice structure, we
will consider quantum state defined on a generic trivalent graph $G$.  The
graph has a branching structure as described by appendix \ref{branching}: Each
edge has an orientation and each vertex has two incoming or one incoming edges.
Each edge has $N+1$ states, labeled by $i=0,...,N$.
Each vertex also has physical states. The number of the states depends on the
states on the connected edges and they are labeled by $\al=1,..., N^{ij}_k$ or
$\bt=1,..., N_{ij}^k$ for   vertices with two incoming and one outgoing edges.
(see Fig. \ref{tri2DC}).

Despite the similar look between $\al$ index and $\u\al$
index, the two indices are very different.  $\u\al$ index
labels the vertices while $\al$ index labels the state on a
vertex. In this paper, we very often use $\al$ to label states on
vertex $\u\al$.

The states on the edge are always bosonic.  However, the states on the vertices
may be fermionic.  We introduce, for example, $F^{ij}_k$ to indecate the number
of fermionic states on the vertex: $\al=1,...,B^{ij}_k$ label the
bosonic(fermion parity even) states and $\al=1+B^{ij}_k,...,F^{ij}_k+B^{ij}_k$ label the fermionic(fermion parity odd)
states.  Here
\begin{align}
 B^{ij}_k=N^{ij}_k-F^{ij}_k
\end{align}
is the number of bosonic states on the vertex.  Similarly, we can introduce
$N_{ij}^k$ and $F_{ij}^k$, to indecate the number of states and fermionic states on vertices with one
incoming edges and two outgoing edges.
In this paper, we will assume that
\begin{align}
 N^{ij}_k  =N_{ij}^k ,\ \ \ \
 B^{ij}_k  =B_{ij}^k ,\ \ \ \
 F^{ij}_k  =F_{ij}^k ,
\end{align}
as required by unitarity.

We introduce $s^{ij}_k(\al)$ to indicate whether a vertex state labeled by $\al$
is bosonic or fermionic: $s^{ij}_k(\al)=0$ if the $\al$-state is bosonic and
$s^{ij}_k(\al)=1$ if the $\al$-state is fermionic.  Here the vertex connects to
three edges $i$, $j$, and $k$ (see Fig. \ref{tri2DC}).
Each graph with a given $\al,\bt,...i,j,...$ labeling (see Fig.
\ref{tri2DC}) corresponds to a state and all such labeled
graphs form an orthonormal basis.  Our fixed-point state is
a superposition of those basis states
\begin{align}
\label{Gstate}
|\psi_\text{fix}\>=\sum_\text{all conf.}
\psi_\text{fix}\left( \bmm \includegraphics[scale=0.18]{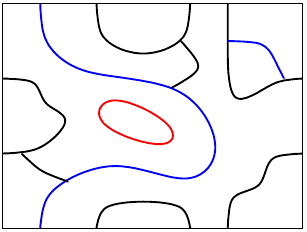}\emm \right)
\left
|\bmm \includegraphics[scale=0.2]{strnet}\emm\right \> .
\end{align}

In the string-net approach, we made a very strong assumption
that the above graphic states on two graphs are the same if
the two graphs have the same topology.  However, since different
vertices and edges are really distinct, a generic graph
state does not have such an topological invariance.  To
consider more general states, in this paper, we would like
to weaken such a topological requirement.  We will consider
vertex-labeled graphs (v-graphs) where each vertex is
assigned an index $\u\al$.  Two v-graphs are said to be
topologically the same if one graph can be continuously
deformed into the other in such a way that vertex labeling
of the two graphs matches.  In this paper, we will consider
the graph states that depend only on the topology of the
v-graphs. Those states are more general than the graph
states that depend only on the topology of the graphs
without vertex labeling.  Such a generalization is
important in obtaining interesting fermionic fixed-point
states on graphs.

\subsection{The structure of a fixed-point
wave function}

\label{entstru}

Before describing the wave function renormalization, we
examine the structure of entanglement of a fixed point
wave function $\psi_\text{fix}$ on a v-graph.
First let us introduce the concept of support space with a
fixed boundary state.

We examine the wave function on a patch,
for example, $\bmm\includegraphics[scale=.35]{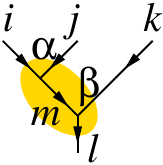}\emm$.
The fixed-point wave function
$\psi_\text{fix}\bpm\includegraphics[scale=.35]{F1gAA}\epm$
(only the relevant part of the graph is drawn)  can be
viewed as a function of $\al,\bt,m$:
$\phi_{ijkl,\Ga}(\al,\bt,m)=\psi_\text{fix} \bpm
\includegraphics[scale=.35]{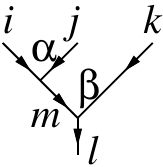} \epm$ if we fix $i,j,k,l$
and the indices on the other part of the graph.  (Here the
indices on the other part of the graph are summarized by $\Ga$.)
As we vary the indices $\Ga$ on the other part of the graph (still
keep $i$, $j$, $k$, and $l$ fixed), the wave function of
$\al,\bt,m$, $\phi_{ijkl,\Ga}(\al,\bt,m)$, may change.  All
those $\phi_{ijkl,\Ga}(\al,\bt,m)$ form a linear space of
dimension $D^{ijk}_{l}$.  $D^{ijk}_{l}$ is an important concept
that will appear later.  We note that the two vertices $\al$
and $\bt$ and the edge $m$ form a region surrounded by the
edges $i,j,k,l$.  So we will call the dimension-$D^{ijk}_{l}$
space the support space $V^{ijk}_{l}$ and $D^{ijk}_{l}$ the support
dimension for the state $\psi_\text{fix}$ on the region
surrounded by a fixed boundary state $i,j,k,l$.

We note that in the fixed-point wave function
$\psi_\text{fix}\bpm\includegraphics[scale=.35]{F1gAA}\epm$,
the number of choices of $\al,\bt,m$ is
$N^{ijk}_{l}=\sum_{m=0}^N N^{ij}_{m} N^{mk}_{l}$.  Thus the
support dimension $D^{ijk}_{l}$ satisfies $D^{ijk}_{l}\leq
N^{ijk}_{l}$.
Here we will make an important assumption -- the
saturation assumption:
\emph{The fixed-point wave
function saturates the inequality:}
\begin{align}
\label{NijklD}
D^{ijk}_{l}= N^{ijk}_{l}\equiv \sum_{m=0}^N N^{ij}_{m} N^{mk}_{l} .
\end{align}
In general, we will make the similar saturation assumption
for any tree graphs.  We will see that the entanglement
structure described by such a saturation assumption is
invariant under the wave function renormalization.

Similarly, we can define $D^{ij}_{k}$ as the support dimension of the
$\Phi_\text{fix}\bpm\includegraphics[scale=.35]{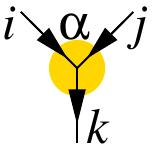}\epm$ on a region
bounded by links $i,j,k$.  Since the region contains only a single vertex
$\u\al$, we have $D^{ij}_k\leq N^{ij}_k$.  The saturation assumption requires
that
\begin{align}
D^{ij}_{k}= N^{ij}_{k} .
\end{align}
In fact, this is how $N^{ij}_{k}$ is defined.

We note that under the saturation assumption, the structure
of the support dimensions for tree graphs is encoded in the
$N^{ij}_{k}$ tensor. Here $N^{ij}_{k}$ plays a similar role as the
pattern of zeros in a classification of fractional quantum
Hall wave functions.\cite{WWsymm}

\subsection{The first type of wave function renormalization: F-move}

Our wave function renormalization scheme contains two types of renormalization.
The first type of renormalization does not change the degrees of freedom and
corresponds to a local unitary transformation.  It corresponds to locally
deforming the v-graph $\bmm \includegraphics[scale=.35]{F1gAA} \emm$ to $\bmm
\includegraphics[scale=.35]{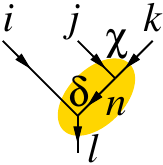} \emm$.  (The parts that are not drawn are
the same.) The fixed-point wave function on the new v-graph is given by
$\psi_\text{fix}\bpm\includegraphics[scale=.35]{F2gAA}\epm$.  Again, such a
wave function can be viewed as a function of $\chi,\del,n$:
$\t\phi_{ijkl,\Ga}(\chi,\del,n)=\psi_\text{fix} \bpm
\includegraphics[scale=.35]{F2gAA} \epm$ if we fix $i,j,k,l$ and the indices on the
other part of the graph.  The support dimension of the state
$\psi_\text{fix}\bpm\includegraphics[scale=.35]{F2gAA}\epm$ on the region
surrounded by $i,j,k,l$ is  $\t D^{ijk}_{l}$.  Again $\t D^{ijk}_{l} \leq \t
N^{ijk}_{l}$, where $\t N^{ijk}_{l}\equiv \sum_{n=0}^N N^{jk}_{n} N^{in}_l$ is the
number of choices of $\chi,\del,n$.  The saturation assumption implies that $
\t N^{ijk}_{l}=\t D^{ijk}_{l} $.

The two fixed-point
wave functions
$\psi_\text{fix}\bpm\includegraphics[scale=.35]{F1gAA}\epm$
and
$\psi_\text{fix}\bpm\includegraphics[scale=.35]{F2gAA}\epm$
are related via a local unitary transformation.
Thus
\begin{align}
 D^{ijk}_{l}=\t D^{ijk}_{l},
\end{align}
which implies
\begin{align}
\label{NNNN}
\sum_{m=0}^N N^{ij}_{m} N^{mk}_{l} =\sum_{n=0}^N N^{jk}_{n} N^{in}_l .
\end{align}
We note that the support space of
$\psi_\text{fix}\bpm\includegraphics[scale=.35]{F1gAA}\epm$
and
$\psi_\text{fix}\bpm\includegraphics[scale=.35]{F2gAA}\epm$
should have the same number of fermionic states.
Thus \eqn{NNNN} can be splitted as
\begin{align}
\label{NNNNb}
 \sum_{m=0}^N
B^{ij}_{m} B^{mk}_{l}
+F^{ij}_{m} F^{mk}_{l}
&=\sum_{n=0}^N
B^{jk}_{n} B^{in}_l
+F^{jk}_{n} F^{in}_l ,
\\
\label{NNNNf}
 \sum_{m=0}^N
B^{ij}_{m} F^{mk}_{l}
+F^{ij}_{m} B^{mk}_{l}
&=\sum_{n=0}^N
B^{jk}_{n} F^{in}_l
+F^{jk}_{n} B^{in}_l .
\end{align}

We express the above unitary transformation
in terms of the tensor
$ F^{ijm,\al\bt}_{kln,\chi\del}$,
where $i,j,k,...=0,...,N$, and
$\al=1,...,N^{ij}_k$, \etc:
\begin{align}
\label{IHwavepsi}
\phi_{ijkl,\Ga}(\al,\bt,m) \simeq
\sum_{n=0}^N
\sum_{\chi=1}^{N^{jk}_{n}}
\sum_{\del=1}^{N^{in}_{l}}
F^{ijm,\al\bt}_{kln,\chi\del}
\t\phi_{ijkl,\Ga}(\chi,\del,n)
\end{align}
or graphically as
\begin{align}
\label{IHwave1}
\psi_\text{fix}
\bpm \includegraphics[scale=.40]{F1g} \epm
\simeq
\sum_{n\chi\del}
F^{ijm,\al\bt}_{kln,\chi\del}
\psi_\text{fix}
\bpm \includegraphics[scale=.40]{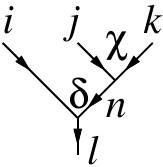} \epm .
\end{align}
where the vertices carrying the states labeled by $(\al,\bt,\chi,\del)$ are
labeled by $(\u\al,\u\bt,\u\chi,\u\del)$ (see Fig.  \ref{tri2DC}).  Here
$\simeq$ means equal up to a constant phase factor.  (Note that the total phase
of the wave function is unphysical.) We will call such a wave function
renormalization step an F-move.

There is a subtlety in \eqn{IHwave1}.
Since some values of $\al$, $\bt$, ... indices
correspond to fermionic states, the sign of wave function
depends on how those fermionic states are ordered.
In \eq{IHwave1}, the wave functions
$ \psi_\text{fix} \bpm \includegraphics[scale=.40]{F1g} \epm$
and
$ \psi_\text{fix} \bpm \includegraphics[scale=.40]{F2g} \epm$
are obtained
by assuming the
fermionic states are order in a particular way:
\begin{align}
&\ \ \ \
\left |\psi_\text{fix} \bpm \includegraphics[scale=.40]{F1g}\epm \right\>
\nonumber\\
&
=\sum \psi^{\u\al\u\bt\u\eta_1\u\eta_2...}_\text{fix} \bpm \includegraphics[scale=.40]{F1g}\epm
|\al\bt\eta_1\eta_2...\>_{\u\al\u\bt\u\eta_1\u\eta_2...}
\nonumber\\
&\ \ \ \
\left |\psi_\text{fix} \bpm \includegraphics[scale=.40]{F2g}\epm \right\>
\nonumber\\
&
=\sum \psi^{\u\chi\u\del\u\eta_1\u\eta_2...}_\text{fix} \bpm \includegraphics[scale=.40]{F2g}\epm
|\chi\del\eta_1\eta_2...\>_{\u\chi\u\del\u\eta_1\u\eta_2...}
\end{align}
where $\sum$ sums over the all indices on the vertices and  
edges. $\eta_i$ are indices on other vertices.  Here
$|\al\bt...\>_{\u\al\u\bt...}$ is a graph state where the
$\u\al$-vertex is in the $|\al\>$-state, the
$\u\bt$-vertex is in the $|\bt\>$-state, \etc.  We note
that $|\bt\al...\>_{\u\bt\u\al...}$ is also a graph state
where the $\u\al$-vertex is in the $|\al\>$-state, the
$\u\bt$-vertex is in the $|\bt\>$-state.  But if the
$|\al\>$-state and the $|\bt\>$-state are fermionic (\ie
${s^{ij}_{m}(\al)=s^{mk}_{l}(\bt)}=1$),
$|\al\bt...\>_{\u\al\u\bt...}$ and
$|\bt\al...\>_{\u\bt\u\al...}$ will differ by a sign, since
in $|\al\bt...\>_{\u\al\u\bt...}$ the fermion on
$\u\bt$-vertex is created before the fermion on
$\u\al$-vertex is created, while
in $|\bt\al...\>_{\u\al\u\bt...}$ the fermion on
$\u\al$-vertex is created before the fermion on
$\u\bt$-vertex is created. In general we have
\begin{align}
 |\al\bt...\>_{\u\al\u\bt...}
=
(-)^{s^{ij}_{m}(\al)s^{mk}_{l}(\bt)}
|\bt\al...\>_{\u\bt\u\al...}.
\end{align}
We see that subscript ${\u\al\u\bt...}$ in
$|\al\bt...\>_{\u\al\u\bt...}$ is important to properly
describe such an order dependent sign.  Similarly, we must
add the superscript in the wave function as well, as in
$\psi^{\u\al\u\bt\u\eta_1\u\eta_2...}_\text{fix} \bpm
\includegraphics[scale=.35]{F1g}\epm$, since the amplitude
of the wave function depends on both the labeled graph $\bmm
\includegraphics[scale=.35]{F1g}\emm$ and the ordering of
the vertices $\u\al\u\bt\u\eta_1\u\eta_2...$.
Such a wave function has the following sign dependence:
\begin{align}
&\ \ \ \
\psi^{\u\al\u\bt\u\eta_1\u\eta_2...}_\text{fix}
\bpm \includegraphics[scale=.40]{F1g}\epm
\nonumber\\
&
=
(-)^{s^{ij}_{m}(\al)s^{mk}_{l}(\bt)}
\psi^{\u\bt\u\al\u\eta_1\u\eta_2...}_\text{fix}
\bpm \includegraphics[scale=.40]{F1g}\epm  .
\end{align}
Thus \eqn{IHwave1} should be more properly written as
\begin{align}
\label{IHwave}
\psi^{\u\al\u\bt...}_\text{fix}
\bpm \includegraphics[scale=.40]{F1g} \epm
\simeq
\sum_{n\chi\del}
F^{ijm,\al\bt}_{kln,\chi\del}
\psi^{\u\chi\u\del...}_\text{fix}
\bpm \includegraphics[scale=.40]{F2g} \epm ,
\end{align}
where the superscripts $\u\al\u\bt...$ and $\u\chi\u\del...$
describing the order of fermionic states are added in the
wave function.

Since the sign of the wave function depends on the ordering
of fermionic states,  the $F$-tensor may also depend on the
ordering. In this paper, we choose a particular ordering of
fermionic states to define the $F$-tensor as described
by $\u\al\u\bt...$ and $\u\chi\u\del...$ in \eqn{IHwave}.
In such a canonical ordering, we create a fermion on the
$\u\bt$-vertex before we create a fermion on the
$\u\al$-vertex.  Similarly, we create a fermion on the
$\u\del$-vertex before we create a fermion on the
$\u\chi$-vertex.

We have seen that, to describe a given fermionic state, the fermionic wave
function $\psi_\text{fix}\left( \bmm \includegraphics[scale=0.18]{strnet}\emm
\right) $ depends on the order of the fermions on the graph.  Different choices
of fermion orders lead to different  fermionic wave functions even for the same
fermionic state.  To avoid such order dependence of the fermionic wave function
(even for the same fermionic state), in the following, we would like to
introduce one Majorana number $\th_{\u\al}$ on each vertex $\u\al$ to rewrite
a wave function that does not depend on the ordering of fermionic states on
vertices.  The Majorana numbers satisfy
\begin{align}
& \th_{\u\al}^2=1,\ \ \ \
\th_{\u\al}\th_{\u\bt}= -\th_{\u\bt} \th_{\u\al} \text{ for any }\al\neq \bt ,
\nonumber\\
& \th_{\u\al}^\dag = \th_{\u\al},\ \ \ \ \
(\th_{\u\al}...\th_{\u\bt})^\dag =
\th_{\u\bt}...\th_{\u\al}.
\end{align}
We introduce the following wave function with Majorana
numbers:
\begin{align}
\Psi_\text{fix} &\bpm \includegraphics[scale=.40]{F1g}\epm
=
[\th_{\u\al}^{s^{ij}_{m}(\al)}\th_{\u\bt}^{s^{mk}_{l}(\bt)}...]
\psi^{\u\al\u\bt...}_\text{fix} \bpm \includegraphics[scale=.40]{F1g}\epm
\nonumber\\
\Psi_\text{fix} &\bpm \includegraphics[scale=.40]{F2g}\epm
=
[
\th_{\u\del}^{s^{in}_{l}(\del)}
\th_{\u\chi}^{s^{jk}_{n}(\chi)}
...]
\psi^{\u\del\u\chi...}_\text{fix} \bpm \includegraphics[scale=.40]{F2g}\epm
\end{align}
where the order of the Majorana numbers $(\th_{\u\al}\th_{\u\bt}...)$ is tied
to the order $\u\al\u\bt...$ in the superscript that describes the order of the
fermionic states.  We see that, by construction, the sign of $\Psi_\text{fix}
\bpm \includegraphics[scale=.35]{F1g}\epm $ does not depend on the order of the
fermionic states, and this is why the Majorana wave function $\Psi_\text{fix}
\bpm \includegraphics[scale=.35]{F1g}\epm $ does not carry the superscript
$(\u\al\u\bt...)$.

We would like to mention that $(\th_{\u\al}, \th_{\u\bt})$ and
$(\th_{\u\chi}, \th_{\u\del})$ are treated as different
Majorana numbers even when, for example, $\u\al$ and $\u\chi$
take the same value.  This is because $\u\al$ and $\u\chi$
label different vertices regardless if $\u\al$ and $\u\chi$
have the same value or not.
So a more accurate notation should be
\begin{align}
\Psi_\text{fix} &\bpm \includegraphics[scale=.40]{F1g}\epm
=
[\th_{\u\al}^{s^{ij}_{m}(\al)}\th_{\u\bt}^{s^{mk}_{l}(\bt)}...]
\psi^{\u\al\u\bt..}_\text{fix} \bpm \includegraphics[scale=.40]{F1g}\epm
\nonumber\\
\Psi_\text{fix} &\bpm \includegraphics[scale=.40]{F2g}\epm
=
[
\t\th_{\u\chi}^{s^{jk}_{n}(\chi)}
\t\th_{\u\del}^{s^{in}_{l}(\del)}
...
]
\psi^{\u\chi\u\del..}_\text{fix} \bpm \includegraphics[scale=.40]{F2g}\epm
,
\end{align}
where $\th_{\u\al}$ and $\t\th_{\u\chi}$ are different even
when $\u\al=\u\chi$.  But in this paper, we will drop the \~{}
and hope that it will not cause any confusions.

Let us introduce the $F$-tensor with Majorana numbers:
\begin{align}
 \cF^{ijm,\al\bt}_{kln,\chi\del}
 =
\th_{\u\al}^{s^{ij}_{m}(\al)}\th_{\u\bt}^{s^{mk}_{l}(\bt)}
\th_{\u\del}^{s^{in}_{l}(\del)} \th_{\u\chi}^{s^{jk}_{n}(\chi)}
F^{ijm,\al\bt}_{kln,\chi\del}
\end{align}
We can rewrite \eq{IHwave} as
\begin{align}
\label{IHwaveG}
 \Psi_\text{fix}
\bpm \includegraphics[scale=.40]{F1g} \epm
\simeq
\sum_{n\chi\del}
 \cF^{ijm,\al\bt}_{kln,\chi\del}
\Psi_\text{fix}
\bpm \includegraphics[scale=.40]{F2g} \epm .
\end{align}
Such an expression is valid for any ordering of the
fermion states.

{}From the graphic representation \eq{IHwave}, We note that
\begin{align}
\label{Feq0}
& F^{ijm,\al\bt}_{kln,\chi\del} = 0 \text{ when}
\\
&
N^{ij}_{m}<1 \text{ or }
N^{mk}_{l}<1 \text{ or }
N^{jk}_{n}<1 \text{ or }
N^{in}_{l}<1
,
\nonumber\\
&
\text{or } s^{ij}_{m}(\al)+s^{mk}_{l}(\bt)
+s^{jk}_{n}(\chi)+s^{in}_{l}(\del)=1 \text{ mod } 2.
\nonumber
\end{align}
When $N^{ij}_{m}<1$ or $ N^{mk}_{l}<1$, the left-hand-side of \eqn{IHwave} is
always zero.  Thus $F^{ijm,\al\bt}_{kln,\chi\del} = 0$ when $N^{ij}_{m}<1$ or $
N^{mk}_{l}<1$.  When $N^{jk}_{n}<1$ or  $N^{in}_{l}<1$, wave function on the
right-hand-side of \eqn{IHwave} is always zero.  So we can choose
$F^{ijm,\al\bt}_{kln,\chi\del} = 0$ when $N^{jk}_{n}<1$ or  $N^{in}_{l}<1$.
Also, $F^{ijm,\al\bt}_{kln,\chi\del}$ represents a pseudo-local bosonic
operator which contains even number of fermionic operators.  Therefore
$F^{ijm,\al\bt}_{kln,\chi\del}$ is non-zero only when
$s^{ij}_{m}(\al)+s^{mk}_{l}(\bt) +s^{jk}_{n}(\chi)+s^{in}_{l}(\del)=0$ mod 2.

For fixed $i$, $j$, $k$, and $l$, the matrix
$F^{ij,\u{\al\bt}}_{kl,\u{\chi\del}}$ with matrix elements
$(F^{ij}_{kl})^{m,\al\bt}_{n,\chi\del} =
F^{ijm,\al\bt}_{kln,\chi\del} $ is a matrix of
dimension $N^{ijk}_{l}$ (see \eq{NNNN}).  Here we require the mapping $
\t\phi_{ijkl,\Ga}(\chi,\del,n) \to \phi_{ijkl,\Ga}(\al,\bt,m)$ generated by the
matrix $F^{ij}_{kl}$ to be unitary.  Since, as we
change $\Ga$, $\t\phi_{ijkl,\Ga}(\chi,\del,n)$ and $\phi_{ijkl,\Ga}(\al,\bt,m)$
span two $N^{ijk}_{l}$ dimensional spaces.  Thus we require
$F^{ij}_{kl}$ to be an $N^{ijk}_{l}\times N^{ijk}_{l}$
unitary matrix
\begin{align}
\label{2FFstar}
\sum_{n\chi\del}
F^{ijm',\al'\bt'}_{kln,\chi\del}
(F^{ijm,\al\bt}_{kln,\chi\del})^*
=\del_{m,m'}\del_{\al,\al'}\del_{\bt,\bt'}.
\end{align}
In this way, the F-move represents an fLU
transformation. It is easy to see that the unitarity condition implies:
\begin{align}
\label{inverseF}
 \Psi_\text{fix}
\bpm \includegraphics[scale=.40]{F2g} \epm
\simeq
\sum_{m\al\bt}
 \left(\cF^{ijm,\al\bt}_{kln,\chi\del}\right)^\dagger
\Psi_\text{fix}
\bpm \includegraphics[scale=.40]{F1g} \epm .
\end{align}

The F-move \eq{IHwaveG} can be viewed as a relationship between
wave functions on different v-graphs that are only differ by a
local transformation.  Since we can locally transform one
v-graph to another v-graph through different paths, the F-move
\eq{IHwaveG} must satisfy certain self consistent
conditions.  For example the v-graph $ \bmm \includegraphics[scale=.35]{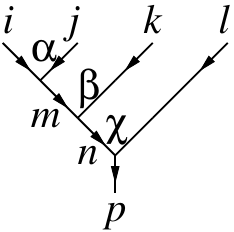}
\emm $ can be transformed to $ \bmm
\includegraphics[scale=.35]{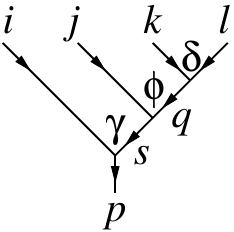} \emm $ through two
different paths; one contains two steps of local
transformations and another contains three steps of local
transformations as described by \eqn{IHwaveG}.  The two
paths lead to the following relations between the wave
functions:
\begin{widetext}
\begin{align}
\label{FFFrelG}
\Psi_\text{fix} \bpm \includegraphics[scale=.40]{pent1g} \epm
&\simeq\sum_{t\eta\vphi}
\cF^{ijm,\al\bt}_{knt,\eta\vphi}
\Psi_\text{fix} \bpm \includegraphics[scale=.40]{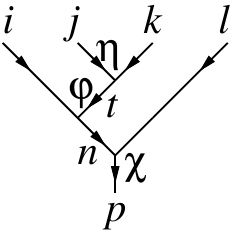} \epm
\simeq\sum_{t\eta\vphi;s\ka\ga}
\cF^{ijm,\al\bt}_{knt,\eta\vphi}
\cF^{itn,\vphi\chi}_{lps,\ka\ga}
\Psi_\text{fix} \bpm \includegraphics[scale=.40]{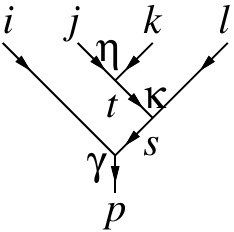} \epm
\nonumber\\
&\simeq\sum_{t\eta\ka;\vphi;s\ka\ga;q\del\phi}
\cF^{ijm,\al\bt}_{knt,\eta\vphi}
\cF^{itn,\vphi\chi}_{lps,\ka\ga}
\cF^{jkt,\eta\ka}_{lsq,\del\phi}
\Psi_\text{fix} \bpm \includegraphics[scale=.40]{pent3g} \epm .
\end{align}
\begin{align}
\label{FFrelG}
\Psi_\text{fix} \bpm \includegraphics[scale=.40]{pent1g} \epm
&\simeq\sum_{q\del\eps}
\cF^{mkn,\bt\chi}_{lpq,\del\eps}
\Psi_\text{fix} \bpm \includegraphics[scale=.40]{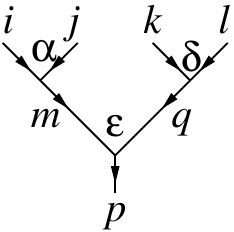} \epm
\simeq\sum_{q\del\eps;s\phi\ga}
\cF^{mkn,\bt\chi}_{lpq,\del\eps}
\cF^{ijm,\al\eps}_{qps,\phi\ga}
\Psi_\text{fix} \bpm \includegraphics[scale=.40]{pent3g} \epm ,
\end{align}
\end{widetext}
The consistence of the above two relations leads
a condition on the $F$-tensor.

To obtain such a condition, let us fix $i$, $j$, $k$, $l$,
$p$, and view $\psi_\text{fix} \bpm \includegraphics[scale=.35]{pent1g}
\epm$ as a function of $\al,\bt,\chi,m,n$:
$\phi(\al,\bt,\chi,m,n)=\psi_\text{fix} \bpm
\includegraphics[scale=.35]{pent1g} \epm$.  As we vary
indices on the other part of graph, we obtain different wave
functions $\phi(\al,\bt,\chi,m,n)$ which form a dimension
$D^{ijkl}_{p}$ space.  In other words, $D^{ijkl}_{p}$ is the
support dimension of the state $\psi_\text{fix}$ on the region
$\al,\bt,\chi,m,n$ with boundary state $i,j,k,l,p$ (see the
discussion in section \ref{entstru}).  Since the number of
choices of $\al,\bt,\chi,m,n$ is $N^{ijkl}_{p}=\sum_{m,n}
N^{ij}_{m} N^{mk}_{n}N^{nl}_{p}$, we have $D^{ijkl}_{p} \leq
N^{ijkl}_{p}$.  Here we require a similar saturation condition as
in \eq{NijklD}:
\begin{align}
\label{NijklpD}
N^{ijkl}_{p} =  D^{ijkl}_{p}
\end{align}
Similarly, the number of choices of $\del,\phi,\ga,q,s$ in
$\psi_\text{fix} \bpm \includegraphics[scale=.35]{pent3g} \epm $ is
also $N^{ijkl}_{p}$. Here we again
assume $\t D^{ijkl}_{p}=N^{ijkl}_{p}$, where
$\t D^{ijkl}_{p}$ is the support dimension of
$\psi_\text{fix} \bpm \includegraphics[scale=.35]{pent3g} \epm $
on the region bounded by $i,j,k,l,p$.

So the two relations \eq{FFrelG} and \eq{FFFrelG}
can be viewed as two relations between a pair of vectors
in the two $D^{ijkl}_{p}$ dimensional vector spaces.
As we vary indices on the other part of graph
(still keep $i,j,k,l,p$ fixed),
each vector in the pair can span the full
$D^{ijkl}_{p}$ dimensional vector space.
So the validity of the two relations \eq{FFrelG} and \eq{FFFrelG}
implies that
\begin{align}
\label{penidG1}
&\ \ \ \
\sum_{t}
\sum_{\eta=1}^{N^{jk}_{t}}
\sum_{\vphi=1}^{N^{it}_{n}}
\sum_{\ka=1}^{N^{tl}_{s}}
\cF^{ijm,\al\bt}_{knt,\eta\vphi}
\cF^{itn,\vphi\chi}_{lps,\ka\ga}
\cF^{jkt,\eta\ka}_{lsq,\del\phi}
\nonumber\\
& \simeq
\sum_{\eps=1}^{N^{mq}_{p}}
\cF^{mkn,\bt\chi}_{lpq,\del\eps}
\cF^{ijm,\al\eps}_{qps,\phi\ga}
.
\end{align}
which is the fermionic generalization of the famous pentagon
identity.  The above expression actually contains many
different pentagon identities, one for each
labeling scheme of the vertices in $\bmm
\includegraphics[scale=.35]{pent1g} \emm$, $\bmm
\includegraphics[scale=.35]{pent2g} \emm$, $\bmm
\includegraphics[scale=.35]{pent3g} \emm$, $\bmm
\includegraphics[scale=.35]{pent4g} \emm$, and $\bmm
\includegraphics[scale=.35]{pent5g} \emm$.
We obtain
\begin{align}
\label{penidG}
&\ \ \ \
\sum_{t}
\sum_{\eta=1}^{N^{jk}_{t}}
\sum_{\vphi=1}^{N^{it}_{n}}
\sum_{\ka=1}^{N^{tl}_{s}}
\cF^{ijm,\al\bt}_{knt,\eta\vphi}
\cF^{itn,\vphi\chi}_{lps,\ka\ga}
\cF^{jkt,\eta\ka}_{lsq,\del\phi}
\nonumber\\
& \simeq
\sum_{\eps=1}^{N^{mq}_{p}}
\cF^{mkn,\bt\chi}_{lpq,\del\eps}
\cF^{ijm,\al\eps}_{qps,\phi\ga}
.
\end{align}
We can use the transformation
\begin{align}
\label{abth}
 F^{ijm,\al\bt}_{kln,\chi\del}
\to \e^{\imth\th} F^{ijm,\al\bt}_{kln,\chi\del}
\end{align}
to change $\simeq$
to $=$ in the above equation and remove the Majorana numbers
to rewrite the above as
\begin{align}
\label{penid}
&\ \ \
\sum_{t}
\sum_{\eta=1}^{N^{jk}_{t}}
\sum_{\vphi=1}^{N^{it}_{n}}
\sum_{\ka=1}^{N^{tl}_{s}}
F^{ijm,\al\bt}_{knt,\eta\vphi}
F^{itn,\vphi\chi}_{lps,\ka\ga}
F^{jkt,\eta\ka}_{lsq,\del\phi}
\nonumber\\
&=
(-)^{s^{ij}_{m}(\al) s^{kl}_{q}(\del)}
\sum_{\eps=1}^{N^{mq}_{p}}
F^{mkn,\bt\chi}_{lpq,\del\eps}
F^{ijm,\al\eps}_{qps,\phi\ga}
.
\end{align}
The above fermionic pentagon identity \eq{penid} is a set of
nonlinear equations satisfied by the rank-10 tensor
$F^{ijm,\al\bt}_{kln,\chi\del}$.  The above consistency
relations
\eq{penid} are
equivalent to the requirement that the local unitary
transformations described by \eqn{IHwaveG} on different
paths all commute with each other up to a total phase
factor.

\subsection{The second type of wave function renormalization: O-move}

The second type of wave function renormalization does change the
degrees of freedom and corresponds to a generalized local
unitary transformation.
One way to implement the second
type renormalization is to reduce
$\bmm
\includegraphics[scale=.35]{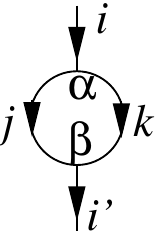} \emm$ to $\bmm
\includegraphics[scale=.35]{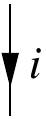} \emm$, so that we still have a trivalent
graph.  This requires that
the support dimension $D_{ii'}$ of
the fixed-point wave function $\psi_\text{fix}
\bpm
\includegraphics[scale=.35]{iOip} \epm $
is given by
\begin{align}
 D_{ii'}=\del_{ii'}.
\end{align}
This implies that
\begin{align}
\psi_\text{fix} \bpm \includegraphics[scale=.40]{iOip} \epm
=\del_{ii'}
\psi_\text{fix} \bpm \includegraphics[scale=.40]{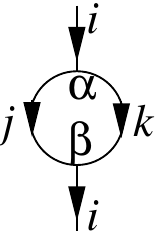} \epm  .
\end{align}
The second type renormalization can now be written as
(since $D_{ii}=1$)
\begin{align}
\label{PhiO}
\psi^{\u\al\u\bt\u\eta...}_\text{fix} \bpm \includegraphics[scale=.40]{iOi} \epm
\simeq O^{jk,\al\bt}_i
\psi^{\u\eta...}_\text{fix} \bpm \includegraphics[scale=.40]{iline} \epm  .
\end{align}
where the ordering of the vertices is described by
$\u\al\u\bt\u\eta...$.
We will call such a wave function renormalization step
an O-move.  Here $O_{i}^{jk,\al\bt} $ satisfies
\begin{align}
\label{Onorm}
\sum_{k,j}
\sum_{\al=1}^{N^{jk}_i}
\sum_{\bt=1}^{N^{jk}_i}
O^{jk,\al\bt}_i (O^{jk,\al\bt}_i)^*=1
\end{align}
and
\begin{align}
\label{Oz}
 O^{jk,\al\bt}_i=0, \text{ if }&
N^{jk}_i<1
\text{ or } s^{jk}_i(\al)+s^{jk}_i(\bt)=1 \text{ mod } 2.
\end{align}
The condition \eq{Onorm} ensures that the two wave functions
on the two sides of \eqn{PhiO} have the same normalization.
We note that the number of choices for the four indices
$(j,k,\al,\bt)$ in $O_{i}^{jk,\al\bt}$ must be equal or
greater than $1$:
\begin{align}
\label{Di1}
 D_i=\sum_{jk} (N^{jk}_i)^2 \geq 1 .
\end{align}
In fact, we should have a stronger condition: the number of
choices for the four indices $(j,k,\al,\bt)$ that correspond
to bosonic states must be equal or greater than $1$
\begin{align}
\label{Di}
 D_i=\sum_{jk} (B^{jk}_i)^2 + (F^{jk}_i)^2
\geq 1 .
\end{align}

The wave functions in \eqn{PhiO} is defined with respect to
the ordering of the fermionic states described by
$\u\al\u\bt\u\eta...$.
Let us introduce
\begin{align}
\Psi_\text{fix} \bpm \includegraphics[scale=.40]{iOi} \epm
&= [
\th_{\u\al}^{s^{jk}_i(\al)}
\th_{\u\bt}^{s^{jk}_i(\bt)}
\th_{\u\eta}^{s(\eta)}...]
\psi^{\u\al\u\bt\u\eta...}_\text{fix} \bpm \includegraphics[scale=.40]{iOi} \epm ,
\nonumber\\
\Psi_\text{fix} \bpm \includegraphics[scale=.40]{iline} \epm
&=[\th_{\u\eta}^{s(\eta)}...]
\psi^{\u\eta...}_\text{fix} \bpm \includegraphics[scale=.40]{iline} \epm
\end{align}
and
\begin{align}
\cO_{i}^{jk,{\al\bt}}
=
\th_{\u\al}^{s^{jk}_i(\al)}
\th_{\u\bt}^{s^{jk}_i(\bt)}
O_{i}^{jk,{\al\bt}} .
\end{align}
We can rewrite \eq{PhiO} as
\begin{align}
\label{PhiOG}
\Psi_\text{fix} \bpm \includegraphics[scale=.40]{iOi} \epm
\simeq \cO_{i}^{jk,{\al\bt}}
\Psi_\text{fix} \bpm \includegraphics[scale=.40]{iline} \epm
\end{align}
which is valid for any ordering of the fermionic states.

\subsection{The third type of wave function renormalization: Y-move}

The third type of wave function renormalization also changes the degrees of
freedom.  The support space of $\bmm \includegraphics[scale=.35]{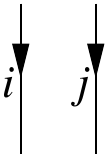} \emm$ is
one dimensional, while the support space of $\bmm
\includegraphics[scale=.35]{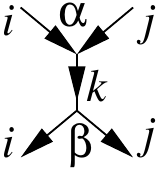} \emm$ is $\sum_k (N^{ij}_k)^2$ dimensional.
So the wave function $\psi_\text{fix}\bpm \includegraphics[scale=.35]{ij}\epm$ is a
particular vector in the support space of $\bmm
\includegraphics[scale=.35]{ijkY} \emm$.
Thus,
the third type of wave function renormalization
takes the following form

\begin{align}
\label{PhiY}
\sum_{k,\al\bt}
Y^{ij}_{k,\al\bt}
\psi_\text{fix}^{\u\al\u\bt\u\eta...}
\bpm \includegraphics[scale=.40]{ijkY} \epm
\simeq
\psi_\text{fix}^{\u\eta...} \bpm \includegraphics[scale=.40]{ij} \epm
\end{align}
where the ordering of the vertices is described by $\u\al\u\bt\u\eta...$.  We
will call such a wave function renormalization step a Y-move.
We can choose
\begin{align}
\label{Yz}
 Y^{ij}_{k,\al\bt}=0, \text{ if }&
N^{ij}_k<1
\text{ or } s^{ij}_k(\al)+s^{ij}_k(\bt)=1 \text{ mod } 2.
\end{align}

The wave functions in \eqn{PhiY} is defined with respect to
the ordering of the fermionic states described by
$\u\al\u\bt\u\eta...$.
Let us introduce
\begin{align}
\Psi_\text{fix} \bpm \includegraphics[scale=.40]{ijkY} \epm
&= [
\th_{\u\al}^{s^{ij}_k(\al)}
\th_{\u\bt}^{s^{ij}_k(\bt)}
\th_{\u\eta}^{s(\eta)}...]
\psi^{\u\al\u\bt\u\eta...}_\text{fix} \bpm \includegraphics[scale=.40]{ijkY} \epm ,
\nonumber\\
\Psi_\text{fix} \bpm \includegraphics[scale=.40]{ij} \epm
&=[\th_{\u\eta}^{s(\eta)}...]
\psi^{\u\eta...}_\text{fix} \bpm \includegraphics[scale=.40]{ij} \epm
\end{align}
and
\begin{align}
\cY_{k,{\al\bt}}^{ij}
=
\th_{\u\bt}^{s^{ij}_k(\bt)}
\th_{\u\al}^{s^{ij}_k(\al)}
Y_{k,{\al\bt}}^{ij} .
\end{align}
We can rewrite \eq{PhiY} as
\begin{align}
\label{PhiYG}
\sum_{k,\al\bt}
\cY_{k,{\al\bt}}^{ij}
\Psi_\text{fix} \bpm \includegraphics[scale=.40]{ijkY} \epm
\simeq
\Psi_\text{fix} \bpm \includegraphics[scale=.40]{ij} \epm
\end{align}
which is valid for any ordering of the fermionic states.

\subsection{A relation between $\cO_{i}^{jk,{\al\bt}}$ and $\cY_{k}^{ij,{\al\bt}}$}

We find that the following wave function has two ways of reduction:
\begin{align}
\label{iOiOOi}
\sum_{\bt\ga}\cY_{i,{\bt\ga}}^{jk}
\Psi_\text{fix} \bpm \includegraphics[scale=.40]{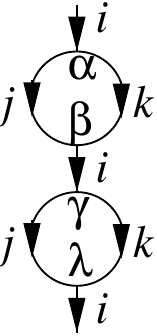} \epm
& \simeq
\Psi_\text{fix} \bpm \includegraphics[scale=.40]{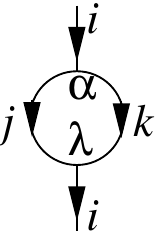} \epm
\nonumber \\
& \simeq
\cO_{i}^{jk,{\al\la}}
\Psi_\text{fix} \bpm \includegraphics[scale=.40]{iline} \epm ,
\end{align}
\begin{align}
&\ \ \ \
\sum_{\bt\ga}\cY_{i,{\bt\ga}}^{jk}
\Psi_\text{fix} \bpm \includegraphics[scale=.40]{iOOi} \epm
\nonumber\\
& \simeq
\sum_{\bt\ga}\cY_{i,{\bt\ga}}^{jk}
\cO_{i}^{jk,{\ga\la}}
\Psi_\text{fix} \bpm \includegraphics[scale=.40]{iOi} \epm
\nonumber \\
& \simeq
\sum_{\bt\ga}\cY_{i,{\bt\ga}}^{jk}
\cO_{i}^{jk,{\ga\la}}
\cO_{i}^{jk,{\al\bt}}
\Psi_\text{fix} \bpm \includegraphics[scale=.40]{iline} \epm
\end{align}
The two reductions should agree, which leads to the condition
\begin{align}
\label{YO1}
\cO_{i}^{jk,{\al\la}}
\simeq
\sum_{\bt\ga}\cY_{i,{\bt\ga}}^{jk}
\cO_{i}^{jk,{\ga\la}}
\cO_{i}^{jk,{\al\bt}}
\end{align}

\subsection{A ``gauge'' freedom}

We note that the following transformation changes the wave function, but does
change fixed-point property and the phase described by the wave function:
\begin{align}
\Psi_\text{fix} \bpm \includegraphics[scale=.40]{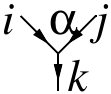} \epm
\to \sum_\bt f^{ij,\al}_{k,\bt} \Psi_\text{fix} \bpm \includegraphics[scale=.40]{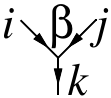} \epm ,
\end{align}
where $f^{ij}_k$ is a unitary matrix
\begin{align}
\sum_\bt f^{ij,\al}_{k,\bt} (f^{ij,\al'}_{k,\bt})^*=\del_{\al\la'} .
\end{align}
Similarly, we have  unitary transformation $f^{k,\al}_{ij,\bt}$ for vertices
with two incoming edges and one outgoing edge.  Such transformations
corrrespond to a choice of basis and should be regarded as an equivalent
relation.

The above transformation induce the following transformation on
$(F^{ijm,\al\bt}_{kln,\ga\la}$,$O^{jk,\al\bt}_i$,$Y_{k,\al\bt}^{ij})$:
\begin{align}
\label{ftrans}
O^{jk,\al\bt}_{i}&\to f^{i,\al}_{jk,\al'} f^{jk,\bt}_{i,\bt'} O^{jk,\al'\bt'}_{i} ,
\nonumber\\
Y^{ij}_{k,\al\bt}&\to (f^{ij,\al'}_{k,\al})^* (f^{k,\bt'}_{ij,\bt})^* Y^{ij}_{k,\al'\bt'} ,
\nonumber\\
 F^{ijm,\al\bt}_{kln,\chi\del} &\to
f^{ij,\al}_{m,\al'}
f^{mk,\bt}_{l,\bt'}
(f^{jk,\chi'}_{n,\chi})^*
(f^{in,\del'}_{l,\del})^*
 F^{ijm,\al'\bt'}_{kln,\chi'\del'} .
\end{align}
We can use the above ``gauge'' degree of freedom to choose
\begin{align}
O^{jk,\al\bt}_{i} = O^{jk,\al}_{i} \del_{\al\bt}, \ \ \ \ O^{jk,\al}_{i} \geq 0.\label{guage}
\end{align}
$O^{jk,\al}_{i}$ is chosen to be a real number.

Then \eqn{YO1} implies that
$ Y^{ij}_{k,\al} \simeq 1/O^{ij,\al}_k$, and we can choose the phase of
$ Y^{ij}_{k,\al} $ to make
\begin{align}
\label{YO}
 Y^{ij}_{k,\al} = 1/O^{ij,\al}_k.
\end{align}

\subsection{Dual F-move and a relation between $O_{i}^{jk,\al}$ and $F^{ijm,\al\bt}_{kln,\del\chi}$}

We also find another wave function that can have two ways of reduction as well:
\begin{align}
&\Psi_\text{fix} \bpm \includegraphics[scale=.40]{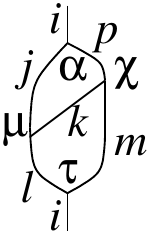} \epm
\simeq
\sum_{s}
\cF^{jkl,\mu\tau}_{mis,\chi\al}
\Psi_\text{fix} \bpm \includegraphics[scale=.40]{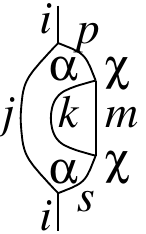} \epm
\nonumber\\
&\simeq
\cF^{jkl,\mu\tau}_{mip,\chi\al}
\cO_{p}^{km,\chi} \cO_{i}^{jp,\al}
\Psi_\text{fix} \bpm \includegraphics[scale=.40]{iline} \epm 
\nonumber \\
&\simeq
\cF^{jkl,\mu\tau}_{mip,\chi\al}
O_{p}^{km,\chi} O_{i}^{jp,\al}(-)^{s_{s}^{km}(\chi)+s_{i}^{jp}(\al)}
\Psi_\text{fix} \bpm \includegraphics[scale=.40]{iline} \epm .
\nonumber
\end{align}
\begin{align}
\Psi_\text{fix} \bpm \includegraphics[scale=.40]{iQQi3} \epm
&\simeq
\t\cF^{jkl,\mu\tau}_{mip,\chi\al}
\Psi_\text{fix} \bpm \includegraphics[scale=.40]{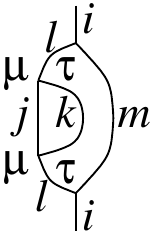} \epm
\nonumber\\
&\simeq
\t\cF^{jkl,\mu\tau}_{mip,\chi\al}
\cO_l^{jk,\mu} \cO_i^{lm,\tau}
\Psi_\text{fix} \bpm \includegraphics[scale=.40]{iline} \epm
\nonumber\\
&\simeq
\t\cF^{jkl,\mu\tau}_{mip,\chi\al}
O_l^{jk,\mu} O_i^{lm,\tau}
\Psi_\text{fix} \bpm \includegraphics[scale=.40]{iline} \epm .
\end{align}
All the edges have a canonical orientation from up to down, and 
$\t \cF^{ijm,\al\bt}_{kln,\chi\del}$, the dual so-called F-move, can be expressed as:
\begin{align}
 \t \cF^{ijm,\mu\tau}_{kln,\chi\al}
 =\th_{\u\al}^{s^{jp}_{i}(\al)}\th_{\u\chi}^{s^{km}_{p}(\chi)}
\th_{\u\mu}^{s^{jk}_{l}(\mu)} \th_{\u\tau}^{s^{lm}_{i}(\tau)}
\t F^{ijm,\mu\tau}_{kln,\chi\al}
\end{align}
This allows us to obtain another condition
\begin{align}
\t F^{jkl,\mu\tau}_{mip,\chi\al}
= F^{jkl,\mu\tau}_{mip,\chi\al}
O_{p}^{km,\chi} O_{i}^{jp,\al}(O_i^{lm,\tau})^{-1}(O_l^{jk,\mu})^{-1}
\end{align}
We require $\t F^{jkl,\mu\tau}_{mip,\chi\al}$ to be unitary, which leads to
\begin{align}
&\ \ \
\sum_{l\mu\tau}
(F^{jkl,\mu\tau}_{mip',\chi'\al'})^*
\frac{O_{p'}^{km,\chi'} O_{i}^{jp',\al'}}{
O_i^{lm,\tau} O_l^{jk,\mu}}
F^{jkl,\mu\tau}_{mip,\chi\al}
\frac{O_{p}^{km,\chi} O_{i}^{jp,\al}}{
O_i^{lm,\tau} O_l^{jk,\mu} }
\nonumber\\
&=\sum_{l\mu\tau}
\left(F^{jkl,\mu\tau}_{mip',\chi'\al'}\right)^*
F^{jkl,\mu\tau}_{mip,\chi\al}
\frac{O_{p'}^{km,\chi'} O_{i}^{jp',\al'}O_{p}^{km,\chi} O_{i}^{jp,\al}
}{
(O_i^{lm,\tau} O_l^{jk,\mu})^2}
\nonumber\\
&=
\del_{pp'}
\del_{\chi\chi'}
\del_{\al\al'}
,
\end{align}
or
\begin{align}
&\sum_{l\mu\tau}
\frac{
(F^{jkl,\mu\tau}_{mip',\chi'\al'})^*
F^{jkl,\mu\tau}_{mip,\chi\al}
}{
(O_i^{lm,\tau} O_l^{jk,\mu})^2}
=
\frac{
\del_{pp'}
\del_{\chi\chi'}
\del_{\al\al'}
}{
(O_{p}^{km,\chi} O_{i}^{jp,\al})^2
}
,
\label{FOcondition}
\end{align}

The above condition can be satisfied by the following ansatz
\begin{align}
 O^{ij,\al}_{k}=\sqrt{\frac{d_id_j}{D d_k}}\del^{ij}_k, \ \ \
D=\sum_l d_l^2,\ \ \ d_i>0,\label{DefO}
\end{align}
where $\delta^{jk}_i=1$ for $N^{jk}_i>0$ and $\delta^{jk}_i=0$ for
$N^{jk}_i=0$.
From \eqn{Onorm}, we find that $d_i$ satisfy
\begin{align}
\label{Nddd}
 \sum_{ij} d_id_j N^{ij}_k = d_kD,\ \ \  D=\sum_l d_l^2.
\end{align}
The solution of such an equation gives us the quantum dimension $d_i$.

\subsection{H-move and an additional constraint between $O_{i}^{jk,\al}$ and $F^{ijm,\al\bt}_{kln,\del\chi}$}
Let us consider a new type of move -- $H$-move.
\begin{align}
\label{Hwave1}
\Psi_\text{fix}
\bpm \includegraphics[scale=.25]{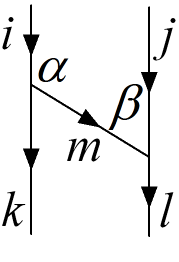} \epm
\simeq
\sum_{n\chi\del}
\mathcal{H}^{kim,\al\bt}_{jln,\chi\del}
\Psi_\text{fix}
\bpm \includegraphics[scale=.25]{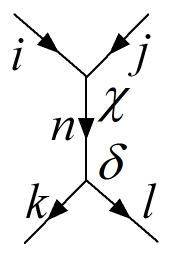} \epm .
\end{align}
Again, we use the convention that all vertices have a canonical ordering from up to down. Similar to the F-move, $\mathcal{H}^{kim,\al\bt}_{jln,\chi\del}$ can be expressed as:
\begin{align}
 \cH^{kim,\al\bt}_{jln,\chi\del}
 =
\th_{\u\al}^{s^{ij}_{m}(\al)}\th_{\u\bt}^{s^{mk}_{l}(\bt)}
\th_{\u\del}^{s^{in}_{l}(\del)} \th_{\u\chi}^{s^{jk}_{n}(\chi)}
H^{kim,\al\bt}_{jln,\chi\del}\label{DefH}
\end{align}
In the following, we will show how to compute the coefficients $H^{kim,\al\bt}_{jln,\chi\del}$ from $F^{ijm,\al\bt}_{kln,\chi\del}$ and $d_i$.

First, by applying the Y-move, we have:
\begin{align}
\label{Hwave2}
\Psi_\text{fix}
\bpm \includegraphics[scale=.25]{H1.png} \epm
\simeq
\sum_{n,\chi^\prime \del}
\mathcal{Y}^{kl}_{n,\chi^\prime \del}
\Psi_\text{fix}
\bpm \includegraphics[scale=.25]{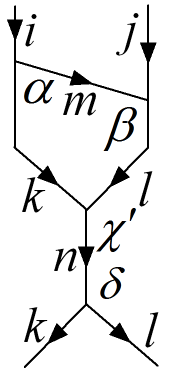} \epm .
\end{align}

Next, by applying an inverse F-move, we obtain:
\begin{align}
\label{Hwave3}
\Psi_\text{fix}
\bpm \includegraphics[scale=.25]{H3.png} \epm
\simeq
\sum_{i^\prime,\beta^\prime\chi}
{\left(\mathcal{F}^{kmi^\prime,\bt^\prime \chi}_{jnl,\bt\chi^\prime}\right)}^\dagger
\Psi_\text{fix}
\bpm \includegraphics[scale=.25]{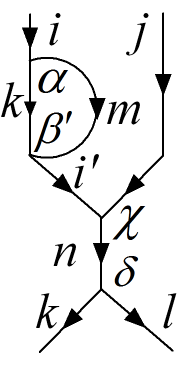} \epm .
\end{align}

Finally, by applying the O-move, we end up with:
\begin{align}
\label{Hwave4}
\Psi_\text{fix}
\bpm \includegraphics[scale=.25]{H4.png} \epm
\simeq
\mathcal{O}^{km,\alpha\beta^\prime}_i\delta_{ii'}
\Psi_\text{fix}
\bpm \includegraphics[scale=.25]{H2.png} \epm .
\end{align}

All together, we find:
\begin{align}
\mathcal{H}^{kim,\al\bt}_{jln,\chi\del}=\sum_{\chi^\prime \beta^\prime}\mathcal{Y}^{kl}_{n,\chi^\prime \del}{\left(\mathcal{F}^{kmi,\bt^\prime \chi}_{jnl,\bt\chi^\prime}\right)}^\dagger \mathcal{O}^{km,\alpha\beta^\prime}_i
\end{align}

Under the proper gauge choice Eq.(\ref{guage}), we can further express the coefficients $H^{ijm,\al\bt}_{kln,\chi\del}$ as:
\begin{align}
 H^{kim,\al\bt}_{jln,\chi\del}&= Y^{kl}_{n,\del}{(F^{kmi,\alpha \chi}_{jnl,\bt\del})}^* O^{km,\alpha }_i\nonumber\\
 &={(F^{kmi,\alpha \chi}_{jnl,\bt\del})}^* {(O^{kl,\del}_{n})}^{-1}O^{km,\alpha }_i
\end{align}
The unitarity condition for H-move requires that:
\begin{align}
\sum_{n\chi\del} \frac{F^{km^\prime i,\alpha \chi}_{jnl,\bt^\prime\del^\prime}{(F^{kmi,\alpha \chi}_{jnl,\bt\del})}^*} {(O^{kl,\del}_{n})^2}=\frac{\delta_{mm^\prime}\delta_{\al\al^\prime}\delta_{\bt\bt^\prime}}{(O^{km,\alpha}_i)^2},
\end{align}

With the special ansatz Eq.(\ref{DefO}), we can further simplify the above expressions as:
\begin{align}
 H^{kim,\al\bt}_{jln,\chi\del}= \sqrt{\frac{d_md_n}{d_id_l}} {\left(F^{kmi,\alpha \chi}_{jnl,\bt\del}\right)}^*
\end{align}
and
\begin{align}
\sum_{n\chi\del} d_n F^{km^\prime i,\alpha \chi}_{jnl,\bt^\prime\del^\prime}{(F^{kmi,\alpha \chi}_{jnl,\bt\del})}^* =\frac{d_id_l}{d_m}
\delta_{mm^\prime}\delta_{\al\al^\prime}\delta_{\bt\bt^\prime},
\end{align}

Similarly, we can also construct the dual-H move:
\begin{align}
\Psi_\text{fix}
\bpm \includegraphics[scale=.25]{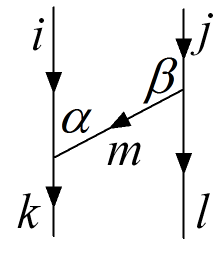} \epm
\simeq
\sum_{n\chi\del}
\t \cH^{kim,\al\bt}_{jln,\chi\del}
\Psi_\text{fix}
\bpm \includegraphics[scale=.25]{H2.png} \epm .
\end{align}
and we can express $\t \cH^{kim,\al\bt}_{jln,\chi\del}$ as:
\begin{align}
\t \cH^{kim,\al\bt}_{jln,\chi\del}
 = \th_{\u\bt}^{s^{mk}_{l}(\bt)}\th_{\u\al}^{s^{ij}_{m}(\al)}
\th_{\u\del}^{s^{in}_{l}(\del)} \th_{\u\chi}^{s^{jk}_{n}(\chi)}
\t H^{kim,\al\bt}_{jln,\chi\del},\label{DefHdual}
\end{align}
where the coefficients $\t \cH^{kim,\al\bt}_{jln,\chi\del}$ can be expressed as:
\begin{align}
\t H^{kim,\al\bt}_{jln,\chi\del}&= Y^{kl}_{n,\del}F^{imk,\alpha \del}_{lnj,\bt\chi} O^{ml,\bt }_j\nonumber\\
 &= F^{imk,\alpha \chi}_{lnj,\bt\del} {(O^{kl,\del}_{n})}^{-1}O^{ml,\bt }_j
\end{align}
Again, with the special ansatz Eq.(\ref{DefO}), we have:
\begin{align}
\t H^{kim,\al\bt}_{jln,\chi\del}=\sqrt{\frac{d_md_n}{d_jd_k}}F^{imk,\alpha \del}_{lnj,\bt\chi}  
\end{align}
It is easy to see that the unitarity condition for dual H-move is automatically satisfied if H-move is unitary.

\subsection{Summary of fixed-point gfLU transformations}

To summarize, the conditions (\ref{NNNN}, \ref{NNNNf},
\ref{Di},
\ref{2FFstar},
\ref{Feq0}, \ref{penid},
\ref{Nddd}
form a set of
non-linear equations whose variables are $N^{ij}_{k}$, $F^{ij}_{k}$,
$F^{ijm,\al\bt}_{kln,\ga\la}$,
$d_i$,
Let us collect those conditions and list them below
\begin{align}
\label{Neq}
&\bullet\
\sum_{m=0}^N N^{ij}_{m} N^{mk}_{l} =\sum_{n=0}^N N^{jk}_{n} N^{in}_l
\nonumber\\
&\bullet\
\sum_{m=0}^N
B^{ij}_{m} F^{mk}_{l}
+F^{ij}_{m} B^{mk}_{l}
=
\sum_{n=0}^N
B^{jk}_{n} F^{in}_l
+F^{jk}_{n} B^{in}_l
\nonumber\\
&\bullet\
\sum_{jk} (B^{jk}_i)^2 + (F^{jk}_i)^2
\geq 1 ,
\nonumber\\
&\bullet\
N^{jk}_i=B^{jk}_i + F^{jk}_i
.
\end{align}
\begin{align}
\label{Feq}
&
\bullet\ \sum_{n\chi\del}
F^{ijm',\al'\bt'}_{kln,\chi\del}
(F^{ijm,\al\bt}_{kln,\chi\del})^*
=\del_{m,m'}\del_{\al,\al'}\del_{\bt,\bt'},
\nonumber\\
&\bullet\
 F^{ijm,\al\bt}_{kln,\chi\del} = 0 \text{ when}
\nonumber \\
&
\ \ \ \
N^{ij}_{m}<1 \text{ or }
N^{mk}_{l}<1 \text{ or }
N^{jk}_{n}<1 \text{ or }
N^{in}_{l}<1
,
\nonumber\\
&
\ \ \ \
\text{or } s^{ij}_{m}(\al)+s^{mk}_{l}(\bt)
+s^{jk}_{n}(\chi)+s^{in}_{l}(\del)=1 \text{ mod } 2.
\nonumber \\
&\bullet\
\sum_{t}
\sum_{\eta=1}^{N^{jk}_{t}}
\sum_{\vphi=1}^{N^{it}_{n}}
\sum_{\ka=1}^{N^{tl}_{s}}
F^{ijm,\al\bt}_{knt,\eta\vphi}
F^{itn,\vphi\chi}_{lps,\ka\ga}
F^{jkt,\eta\ka}_{lsq,\del\phi}
\nonumber\\
&=
(-)^{s^{ij}_{m}(\al) s^{kl}_{q}(\del)}
\sum_{\eps=1}^{N^{mq}_{p}}
F^{mkn,\bt\chi}_{lpq,\del\eps}
F^{ijm,\al\eps}_{qps,\phi\ga}
.
\end{align}
\begin{align}
\label{Oeq}
\bullet\  \sum_{i,j} d_id_j N^{ij}_k =d_kD,\ \ \  D=\sum_l d_l^2.
\end{align}
\begin{align}
\label{dFeq}
\bullet\  \sum_{n\chi\del} d_n F^{km^\prime i,\alpha \chi}_{jnl,\bt^\prime\del^\prime}{(F^{kmi,\alpha \chi}_{jnl,\bt\del})}^* =\frac{d_id_l}{d_m}
\delta_{mm^\prime}\delta_{\al\al^\prime}\delta_{\bt\bt^\prime},
\end{align}
Finding $N^{ij}_{k}$, $F^{ij}_{k}$, $F^{ijm,\al\bt}_{kln,\ga\la}$,
and $d_i$ that satisfy such a set of
non-linear equations corresponds to finding a fixed-point gfLU transformation
that has a non-trivial fixed-point wave function.  So the solutions
$(N^{ij}_{k}$,$F^{ij}_{k}$,$F^{ijm,\al\bt}_{kln,\ga\la}$,$d_i)$
give us a characterization of fermionic topological orders (and bosonic
topological orders as a special case where $F^{ij}_{k}=0$).

We would like to stress that, although the solutions
$(N^{ij}_{k}$,$F^{ij}_{k}$,$F^{ijm,\al\bt}_{kln,\ga\la}$,$d_i)$ decscribe 2+1D
fermionic topological orders with gappable edge, the correspondence is not one
to one.  Given a set of solutions, the transformation in \eqn{ftrans} on
$F^{ijm,\al\bt}_{kln,\ga\la}$ will generate another set of solutions (since the
equitions for $d_i$ and $F^{ijm,\al\bt}_{kln,\ga\la}$ decouple).  The two sets
of solutions describe the same topologically ordered phase. Also \eqn{ftrans}
does not include all the redundancy: two solutions that are not related by
``gauge'' transformation \eqn{ftrans} may still describe the same fermionic
topological orders.  We need to compute the modula transformation $T$ and $S$
from the data $(N^{ij}_{k}$,$F^{ij}_{k}$,$F^{ijm,\al\bt}_{kln,\ga\la}$,$d_i)$
to determine the 2+1D topological order.\cite{Wrig,KW9327,LW1384}

\section{Categorical framework}

To provide a conceptual understanding of our generalization
of string-net model, we discuss briefly the categorical
picture which underlies earlier algebraic manipulations.
Such a mathematical framework will provide more examples for
our fermionic string-net Hamiltonians in Appendix \ref{fHam}.

A string-net or Levin-Wen Hamiltonian can be easily
constructed using $6j$-symbols from a unitary fusion
category $\calC$.  The elementary excitations of the model
form a unitary modular tensor category (UMTC) $\cal{E}$,
which turns out to be the quantum double $Z(\calC)$  of the
input category $\calC$.  A priori, the output modular category
$\cal{E}$ is not necessarily related to the input category
$\calC$.  Therefore, it is conceivable that similar
Hamiltonians can be constructed from some other algebraic
data where the elementary excitations still form a UMTC,
which is not necessarily a quantum double.  This is explored
in \Ref{CGW1038}.  In the preceding sections, we generalize
the string-net model by including fermionic degrees of
freedom.

The mathematical framework for such a generalization is the
theory of enriched categories.\cite{K051}  An enriched category is
actually not a category, just like a quantum group is not a
group.  We will consider only special enriched categories,
which we call {\it projective super fusion categories.}  The
ordinary unitary fusion categories are enriched categories
over the category of Hilbert spaces, while projective super
fusion categories are enriched categories over the category
of super Hilbert spaces up to {\it projective even} unitary
transformations.

To the physically inclined readers, the use of category
theory in condensed matter physics seems to be unjustifiably
abstract.  We would argue that the abstractness of category
theory is actually its virtue.  Topological properties of
quantum systems are independent of the microscopic details
and are non-local.  A framework to encode such properties is
necessarily blind to microscopic specifics.  Therefore,
philosophically category theory could be extremely relevant,
as we believe.

\subsection{Projective super tensor category}

We use super vector spaces to accommodate fermionic states,
and generalize the composition of linear transformations to
one only up to overall phases---a possibility allowed by
quantum mechanics.  The projective tensor category of vector
spaces is the category of vector spaces and linear
transformations composed up to overall phases, and the
category of super vector spaces is the tensor category of
$\Z_2$-graded vector spaces and all {\it even} linear
transformations.

In the categorical language, a fusion category is a rigid
finite linear category with a simple unit.  Equivalently, it
can be defined using $6j$-symbols: an equivalence class of
solutions of pentagons satisfying certain normalizations\cite{Wang10}.
Fermionic $6j$-symbols $\cF^{ijm,\al\bt,a}_{kln,\ga\la,b}$ in
\eqn{IHwaveG} with certain normalizations define a
projective super fusion category if they satisfy fermionic
pentagon equations \eqn{penidG1}.  However, the setup used
in this paper may only generate a subclass of projective
super fusion category.

\subsection{Super tensor category from super quantum groups}

The trivial example of a super tensor category is the
category of $\Z_2$-graded vector spaces and all linear
transformations.  More interesting examples of super tensor
categories can be constructed from the representation theory
of super quantum groups.

Super quantum groups are deformations of Lie
superalgebras.\cite{CK9072,MT9457} Though a mathematical theory
analogous to quantum group exists, the details have not been worked
out enough for our application here. In literature, the categorical
formulation focuses on the invariant spaces of even entwiners, while
for our purpose, we need to consider all entwiners. In particular,
we are not aware of work on Majorana valued Clebsch-Gordon
coefficients, therefore, we will leave the details to future
publications.

\section{Simple solutions of the fixed-point conditions}

In this section, let us discuss some simple solutions of the fixed-point
conditions (\ref{Neq}, \ref{Feq}, \ref{Oeq}, \ref{dFeq}) for the fixed-point
gfLU transformations $(N^{ij}_{k},F^{ij}_{k},F^{ijm,\al\bt}_{kln,\ga\la},
d_i)$.

\subsection{Solutions from group cohomology}

Many bosonic solutions can be constructed from a finite group $G$.  Here we
treat the edge index $i,j,k,\cdots$ as elements in the group: $i,j,k \in G$
with group mulitplication $i\cdot j \in G$.  We choose the fusion coefficient as
\begin{align}
 N^{ij}_k &=\begin{cases}
1, &\text{ if } i\cdot j=k\\
0, &\text{ if } i\cdot j\neq k\\
\end{cases}
\nonumber\\
F^{ij}_k &= 0.
\end{align}
Since $N^{ij}_k=0,1$, we can drop the indices $\al,\bt$ on vertices.
$F^{ijm,\al\bt}_{kln,\ga\la}$ that satisfies \eqn{Feq} is given by
\begin{align}
 F^{ijm}_{kln}=\om_3(i,j,k)
N^{ij}_m
N^{mk}_l
N^{jk}_n
N^{in}_l ,
\end{align}
where $\om_3(i,j,k)$ is the 3-cocycle in group cohomology
class $\cH^3[G,U(1)]$.\cite{CGL1172,CGL1204}
In this case, the self-consistent condition Eq.(\ref{Feq}) for $F$-tensor becomes the cocycle equation for $\om_3(i,j,k)$:
\begin{align}
\om_3(i,j,k)\om_3(i,j\cdot k,l)\om_3(j,k,l)=\om_3(i\cdot j,k,l)\om_3(i,j,k \cdot l)
\end{align}
$d_i$ that satisfies \eqn{Oeq} is given by
\begin{align}
 d_i=1.
\end{align}
Such a solution describes a ``twisted'' gauge theory in
2+1D.\cite{DW9093,HW1232,HWW1314} If we choose a trivial 3-cocycle
$\om_3(i,j,k)=1$, the solution will describe a standard gauge theory with gauge
group $G$ in 2+1D.

\subsection{Solutions from group supercohomology}
Similarly, many fermionic solutions can be constructed from a finite group $G$.  Again we
treat the edge index $i,j,k,\cdots$ as elements in the group: $i,j,k \in G$
with group mulitplication $i\cdot j \in G$.  We choose the same fusion coefficient $N^{ij}_k$ as for bosonic solutions, but with
nonzero $F^{ij}_k$: 
\begin{align}
 N^{ij}_k &=\begin{cases}
1, &\text{ if } i\cdot j=k\\
0, &\text{ if } i\cdot j\neq k\\
\end{cases}
\nonumber\\
F^{ij}_k &=n_2(i,j)N^{ij}_k\neq 0.
\end{align}
where $n_2(i,j)\in \mathbb{Z}_2$ valued on $0,1$ is 2-cocycle in the obstruction free subgroup of group cohomology
class $B\cH^2[G,\mathbb{Z}_2]$. By obstruction free, we mean that for any $n_2(i,j)$ satisfying the 
2-cocycle condition:
\begin{align}
n_2(i,j)+n_2(i\cdot j, k)=n_2(i, j\cdot k)+n_2(j, k),
\end{align}
the following $\pm 1$ valued function:
\begin{align}
(-)^{n_2(i,j)n_2(k,l)},
\end{align}
must be a coboundary in $\mathcal{B}^4[G,U(1)]$ when we view it as a 4-cocycle
with $U(1)$ coefficient. Each element in $B\cH^2[G,\mathbb{Z}_2]$ become a
valid $Z_2$-graded structure for fermion systems.

On the other hand, since $N^{ij}_k=0,1$, we can again drop the indices $\al,\bt$ on vertices.
$F^{ijm,\al\bt}_{kln,\ga\la}$ that satisfies \eqn{Feq} is given by
\begin{align}
 F^{ijm}_{kln}=\t\om_3(i,j,k)
N^{ij}_m
N^{mk}_l
N^{jk}_n
N^{in}_l ,
\end{align}
where $\om_3(i,j,k)$ is the 3-supercocycle in group supercohomology
class $\cH^3_f[G,U(1)]$\cite{CGL1172,CGL1204}, which satisfies:
\begin{align}
&\t\om_3(i,j,k)\t\om_3(i,j\cdot k,l)\t\om_3(j,k,l)\nonumber\\
&=(-)^{n_2(i,j)n_2(k,l)}\t\om_3(i\cdot j,k,l)\t\om_3(i,j,k \cdot l)
\end{align}
Again $d_i$ that satisfies \eqn{Oeq} is given by
\begin{align}
 d_i=1.
\end{align}
Such a solution describe a fermionic gauge theory in
2+1D, e.g. the recently proposed fermionic toric code.\cite{GWW1332}

\section{Summary}

Using string-net condensations and LU transformations (or in other words,
unitary fusion category theory), we have obtained a classification of 2+1D
topological orders with gappable edge in bosonic
systems.\cite{LWstrnet,CGW1038} An interacting fermionic system is a non-local
bosonic system. So classifying topological orders in fermion systems appears to
be a very difficult problem.

In this paper, we introduced fLU and gfLU transformations, which allow us to
develop a general theory for a large class of fermionic topological orders.  We
propose that 2+1D topological orders with gappable edge in fermionic systems
can be classified by the data $( N^{ij}_k, F^{ij}_k,
F^{ijm,\al\bt}_{jkn,\chi\del},d_i)$ that satisfy a set of non-linear algebraic
equations \eq{Neq}, \eq{Feq}, \eq{Oeq}, and \eq{dFeq}.  Such a result generalizes the
string-net result\cite{LWstrnet,CGW1038} to fermionic cases.  We hope our
approach to be a starting point for establishing a mathematical framework for
topological orders in interacting fermion systems.

We would like to thank M. Levin for some very helpful discussions.  XGW is
supported by NSF Grant No.  DMR-1005541 and NSFC 11274192.  He is also
supported by the BMO Financial Group and the John Templeton Foundation  Grant
No. 39901.  ZCG is supported in part by the NSF Grant No.  NSFPHY05-51164.
Research at Perimeter Institute is supported by the Government of Canada
through Industry Canada and by the Province of Ontario through the Ministry of
Research.

\appendix

\begin{figure}[tb]
\begin{center}
\includegraphics[scale=0.6]{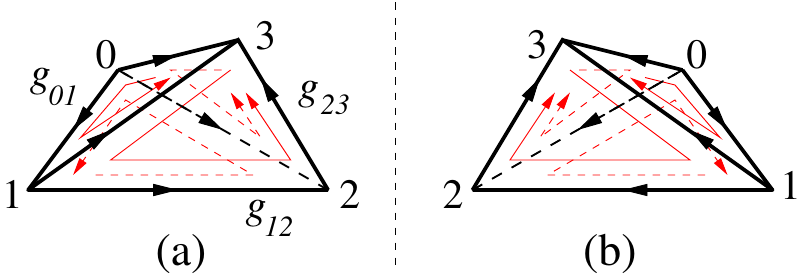} \end{center}
\caption{ (Color online) Two branched simplices with opposite orientations.
(a) A branched simplex with positive orientation and (b) a branched simplex
with negative orientation.  }
\label{mir}
\end{figure}

\section{Branching structure of 2D graph}
\label{branching}

To define a lattice model a space $M$, we first triangulate of the space $M$ to
obtain a complex $M_\text{tri}$.  We will call a cell in the complex as a
simplex.  In order to define a generic lattice theory on the complex
$M_\text{tri}$, it is important to give the  vertices of each simplex a local
order.  A nice local scheme to order  the  vertices is given by a branching
structure.\cite{C0527,CGL1172,CGL1204} A branching structure is a choice of
orientation of each edge in the $n$-dimensional complex so that there is no
oriented loop on any triangle (see Fig. \ref{mir}).

The branching structure induces a \emph{local order} of the vertices on each
simplex.  The first vertex of a simplex is the vertex with no incoming edges,
and the second vertex is the vertex with only one incoming edge, \etc.  So the
simplex in  Fig. \ref{mir}a has the following vertex ordering: $0,1,2,3$.

The branching structure also gives the simplex (and its sub simplexes) an
orientation denoted by $s_{ij \cdots k}=\pm$.  Fig. \ref{mir} illustrates two
$3$-simplices with opposite orientations $s_{0123}=+$ and $s_{0123}=-$.  The
red arrows indicate the orientations of the $2$-simplices which are the
subsimplices of the $3$-simplices.  The black arrows on the edges indicate the
orientations of the $1$-simplices.

In this paper, we will only consider 2D space.  The graph that we use to define
our lattice model is the dual graph of the 2D complex $M_\text{tri}$. The
branching structure of $M_\text{tri}$ leads a branching structure of our graph:
each vertex of the graph cannot have three incoming edges or three outgoing
edges.

\section{The parent Hamiltonian for fixed point wavefunctions}
\subsection{The fermionic structure of support space}
\label{fspsp}

To understand the fermionic structure of the support space
 $\t V_A$, let us first study the
structure of $\rho_A$.  Let $|\phi_{i}\>$ be a basis of the
Hilbert space of the region $A$ and $|\bar\phi_{\bar i}\>$
be a basis of the Hilbert space of the region outside of
$A$.  $|\psi\>$ can be expanded by $|\phi_{i}\>\otimes
|\bar\phi_{\bar i}\>$:
\begin{align}
 |\psi\>=\sum_{i,\bar i} C_{i,\bar i} |\phi_{i}\>\otimes
|\bar\phi_{\bar i}\>.
\end{align}
Then the matrix elements of $\rho_A$ is given by
\begin{align}
 (\rho_A)_{ij} =\sum_{\bar i} (C_{i\bar i})^* C_{j\bar i} .
\end{align}

For a fermion system, the Hilbert space on a site, $V_i$
has a structure: $V_i=V_i^0\oplus V_i^1$, where states in
$V_i^0$ have even numbers of fermions and states in $V_i^1$
have odd numbers of fermions.  The Hilbert space on the
region $A$, $V_A$, has a similar structure $V_A=V_A^0\oplus
V_A^1$, where states in $V_A^0$ have even number of fermions
and states in $V_A^1$ have odd numbers of fermions.  Let
$|\phi_{i,\al}\>$ be a basis of $V_A^\al$.  Similarly, the
Hilbert space on the region outside of $A$, $V_{\bar A}$,
also has a structure $V_{\bar A}=V_{\bar A}^0\oplus V_{\bar
A}^1$.  Let $|\bar\phi_{\bar i,\al}\>$ be a basis of $V_{\bar
A}^\al$.  In this case, $|\psi\>$ can be expanded as
\begin{align}
 |\psi\>=\sum_{i\al;\bar i\bt} C_{i,\al;\bar i,\bt} |\phi_{i,\al}\>\otimes
|\bar\phi_{\bar i,\bt}\>.
\end{align}
the matrix elements of $\rho_A$ can now be expressed as
\begin{align}
 (\rho_A)_{i,\al;j,\bt} =\sum_{\bar i,\ga}
(C_{i,\al;\bar i,\ga})^* C_{j,\bt;\bar i,\ga} .
\end{align}
Since the fermion number mod 2 is conserved, we may assume
that $|\psi\>$ contains even numbers of fermions.  This
means $C_{i,\al;\bar i,\ga}=0$ when $\al+\ga=1$ mod 2.
Hence, we find that
\begin{align}
 (\rho_A)_{i,\al;j,\bt} =0, \text{ when } \al+\bt =1 \text{
mod } 2.
\end{align}
Such a density matrix tells us that the support space $\t
V_A$ has a structure $ \t V_A = \t V_A^0 \oplus \t V_A^1 $,
where $\t V_A^0$ has even numbers of fermions and $\t V_A^1$
has odd numbers of fermions.  This means that $U_g$ contains
only even numbers of fermionic operators (\ie $U_g$ is
a pseudo-local bosonic operator).

\subsection{Compute the parent Hamiltonian}
\label{fHam}

\begin{figure}
\begin{center}
\includegraphics[scale=0.45]{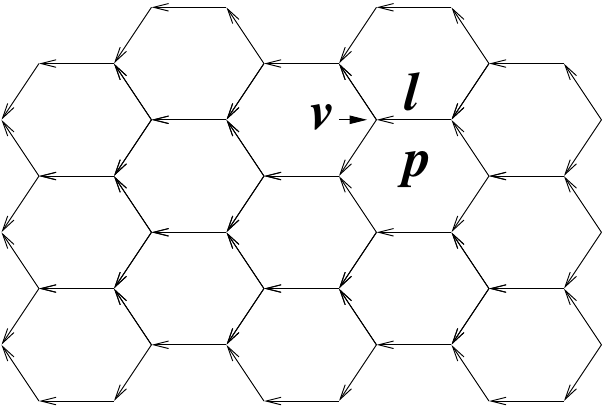}
\end{center}
\caption{
A honeycomb lattice. The vertices are labeled by $\v v$,
hexagons by $\v p$, and links by $\v l$.
}
\label{Hlatt}
\end{figure}

In the section \ref{fixwv}, we have constructed the fixed-point
wave functions from the solutions $(N^{ij}_k, F^{ij}_k,
F^{ijm,\al\bt,a}_{kln,\ga\la,b},d_i)$
of the self consistent conditions.  In this section, we will
show that those fixed-point wave functions on a honeycomb
lattice (see Fig. \ref{Hlatt}) are exact gapped ground state
of a local Hamiltonian
\begin{align}
 \hat H=\sum_{\v v} (1-\hat Q_{\v v}) + \sum_{\v p} (1-\hat B_{\v p})
\end{align}
where $\sum_{\v v}$ sums over all vertices and
$\sum_{\v p}$ sums over all hexagons.
 
The Hamiltonian $\hat H$ should act on the Hilbert space
$V_G$ formed by all the graph states.  It turns out that it
is more convenient to write down the Hamiltonian if we
expand the Hilbert space by adding an auxiliary qubit to each vertex:
\begin{align}
 V^{ex}_G=V_G\otimes (\otimes_{\v v} V_{qubit})
\end{align}
where $\otimes_{\v v}$ goes over all vertices and
$V_{qubit}$ is the two dimensional Hilbert space of an auxiliary qubit
$|I\>$, $I=0,1$.  So in the expanded Hilbert space
$V^{ex}_G$, the states on a vertex $\v v$ are labeled by
$|\al\>\otimes|I\>$, $I=0,1$.  $V_G$ is embedded into
$V^{ex}_G$ in the following way: each vertex state $|\al\>$
in $V_G$ correspond to the following vertex state
$|\al\>\otimes |s_{ijk}(\al)\>$ in $V^{ex}_G$, where we have
assume that the states on the three links connecting to the
vertex are $|i\>$, $|j\>$, and $|k\>$.  So the new auxiliary qubit
$|I\>$ on a vertex is completely determined by $(i,j,k,\al)$
and does not represent an independent degree of freedom. It
just tracks if the vertex state is bosonic or fermionic.
The $|0\>$-state correspond to bosonic vertex states and the
$|1\>$-state correspond to fermionic vertex states.

In the expanded Hilbert space, $\hat Q_{\v v}$ in
$\hat H$ acts on the
states on the 3 links that connect to the vertex $\v v$ and on
the states $|\al\>\otimes|I\>$ on the vertex $\v v$:
\begin{align}
& \hat Q_{\v v} \left | \bmm \includegraphics[scale=.40]{ijk} \emm  \right \>
\otimes|I\>
=
\left | \bmm \includegraphics[scale=.40]{ijk} \emm  \right \>
\otimes|I\>
\ \text{ if } N^{ij}_k>0,\ I=s^{ij}_k(\al),
\nonumber\\
& \hat Q_{\v v} \left | \bmm \includegraphics[scale=.40]{ijk} \emm  \right \>
\otimes|I\>
=0
\ \ \ \ \text{ otherwise} .
\end{align}
Clearly, $\hat Q_{\v v}$ is a projector $\hat Q_{\v
v}^2=\hat Q_{\v v}$.  The $\hat B_{\v p}$ operator in $\hat
H$ acts on the states on the 6 links and the 6 vertices of
the hexagon $\v p$ and on the 6 links that connect to the
hexagon.  However, $\hat B_{\v p}$ operator will not alter
the states on the 6 links that connect to the hexagon. Let us define the Majorana number valued matrix element 
$\cB^{a\al,b\bt,c\ga,d\la,e\mu,f\nu } _{a'\al',b'\bt',c'\ga',d'\la',e'\mu',f'\nu' } (i,j,k,l,m,n)$ as:
\begin{align}
&\cB^{a\al,b\bt,c\ga,d\la,e\mu,f\nu } _{a'\al',b'\bt',c'\ga',d'\la',e'\mu',f'\nu' } (i,j,k,l,m,n)\nonumber\\ &=
\left\langle\Psi_\text{fix} \bpm \includegraphics[scale=.25]{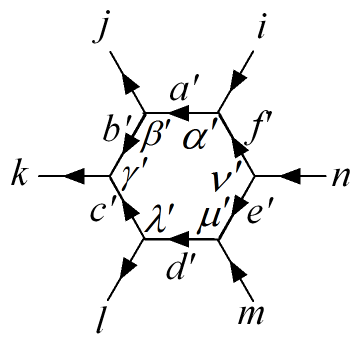} \epm \right|
\hat B_{\v p}
\left|\Psi_\text{fix} \bpm \includegraphics[scale=.25]{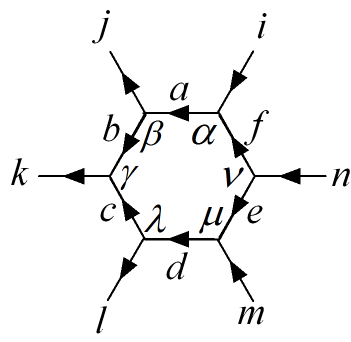} \epm \right\rangle
\end{align}

In order to compute $\cB$, we consider the following local deformation: 
\begin{align}
\Psi_\text{fix} \bpm \includegraphics[scale=.25]{Ha1} \epm
\to
\Psi_\text{fix} \bpm \includegraphics[scale=.25]{Ha7} \epm
\end{align}
We note that the
self consistent conditions satisfied by the $F$-tensor and
$O$-tensor ensure that all those different ways to transform
between the two above states
lead to the same $\cB$ matrix. 

To understand how $\cB$ acts on a state that is not in the
support space, let us consider the dimension $D_{ijklmn}$ of
the support space $V_{ijklmn}$ which can be calculated by
deforming the graph $ \bmm
\includegraphics[scale=.25]{Ha1} \emm$ into $ \bmm
\includegraphics[scale=.25]{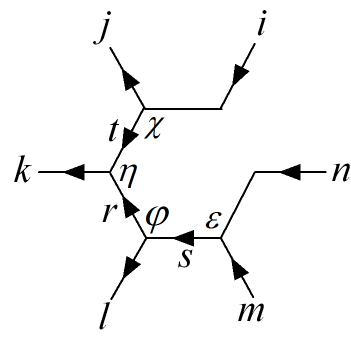} \emm$ through a gfLU
transformation $\cU$.  Under the saturation assumption,
$D_{ijklmn}$ is equal to the distinct labels in the graph $
\bmm \includegraphics[scale=.25]{Ha6} \emm$ (with
$i,j,k,l,m,n$ fixed):
\begin{align}
D_{ijklmn} = \sum_{trs} N^{jt}_i N^{tr}_k N^{rl}_s N^{nm}_s .
\end{align}

\begin{figure}[tb]
\begin{center}
\includegraphics[scale=0.35]{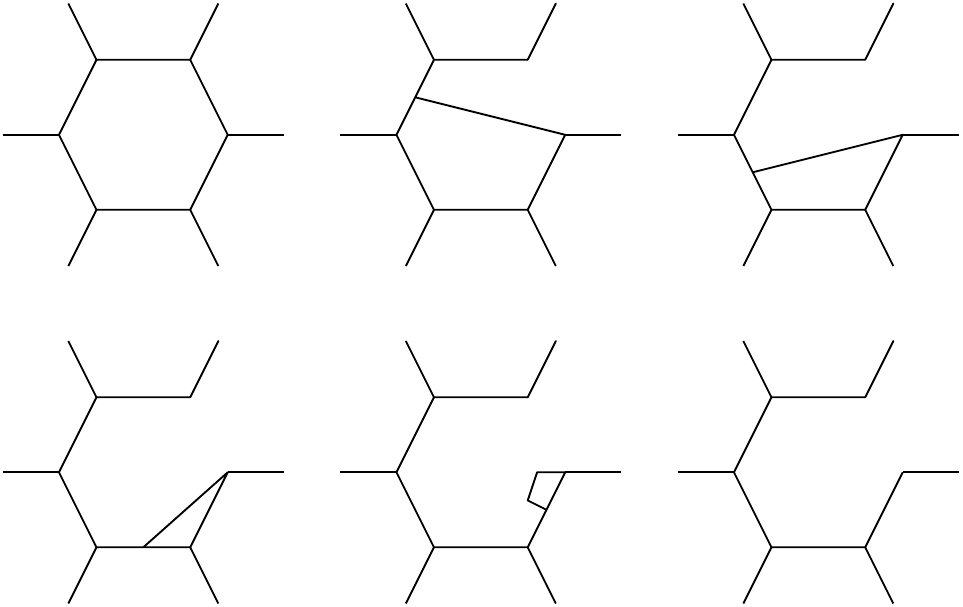}
\end{center}
\caption{
$\cU_P$ is generated by an inverse H-move, an F-move, a dual H-move, an inverse F-moves and finally one O-move, which
turns a hexagon graph into a tree graph.
$(\cU_P)^\dag$ turns a tree graph into a hexagon graph.
}
\label{hextotree}
\end{figure}

In particular, we
can have the following two paths that connect the two states
(see Fig. \ref{hextotree}):
\begin{align*}
\begin{CD}
\Psi_\text{fix} \bpm \includegraphics[scale=.25]{Ha1} \epm
@>\cB>>
\Psi_\text{fix} \bpm \includegraphics[scale=.25]{Ha7} \epm
\\
@VV\cU_P V
@AA\cU_P^\dag A \\
\Psi_\text{fix} \bpm \includegraphics[scale=.25]{Ha6} \epm
@>\cC>>
\Psi_\text{fix} \bpm \includegraphics[scale=.25]{Ha6} \epm
\end{CD}
\end{align*}
where $\cU$ is a gfLU transformation.  We find that
\begin{align}
 \cB = \cU_P^\dag \cC \cU_P
\end{align}
where $\cC$, acting on $\Psi_\text{fix} \bpm
\includegraphics[scale=.25]{Ha6} \epm $, is a dimension
$D_{ijklmn}\times D_{ijklmn}$ identity matrix. In the following let us compute the explicit form of $\cU_P$.

As seen in Fig. \ref{hextotree}, let us first apply an inverse H-move, an F-move, a dual H-move, an inverse F-moves and finally one O-move, thus we obtain:
\begin{widetext}
\begin{eqnarray}
\Psi_\text{fix} \bpm \includegraphics[scale=.25]{Ha1} \epm &=&\sum_{t\chi\delta}(\cH^{jit,\chi\delta}_{fba,\al\bt})^\dagger
\Psi_\text{fix} \bpm \includegraphics[scale=.25]{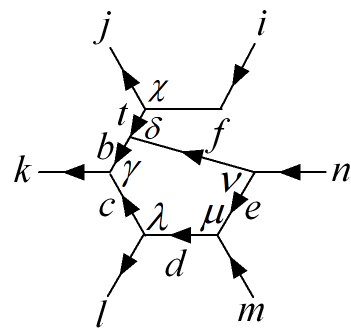} \epm \\
&=& \sum_{t\chi\delta}\sum_{r\kappa\eta}(\cH^{jit,\chi\delta}_{fba,\al\bt})^\dagger \cF^{tfb,\delta\gamma}_{ckr,\kappa\eta}
\Psi_\text{fix} \bpm \includegraphics[scale=.25]{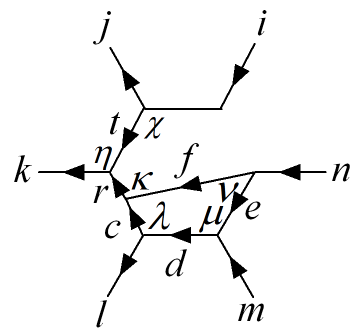} \epm \nonumber\\
&=& \sum_{t\chi\delta}\sum_{r\kappa\eta}\sum_{s\rho\varphi}(\cH^{jit,\chi\delta}_{fba,\al\bt})^\dagger \cF^{tfb,\delta\gamma}_{ckr,\kappa\eta}\t \cH^{rfc,\kappa\lambda}_{dls,\rho\varphi}
\Psi_\text{fix} \bpm \includegraphics[scale=.25]{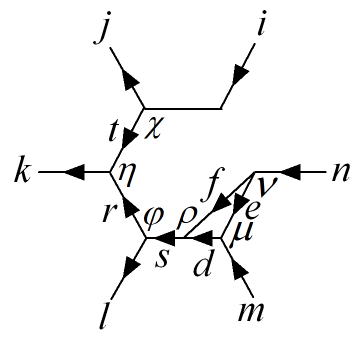} \epm \nonumber\\
&=& \sum_{t\chi\delta}\sum_{r\kappa\eta}\sum_{s\rho\varphi}\sum_{n^\prime\nu^\prime\epsilon}(\cH^{jit,\chi\delta}_{fba,\al\bt})^\dagger \cF^{tfb,\delta\gamma}_{ckr,\kappa\eta}\t \cH^{rfc,\kappa\lambda}_{dls,\rho\varphi}
(\cF^{fen^\prime,\nu^\prime\epsilon}_{msd,\mu\rho})^\dagger
\Psi_\text{fix} \bpm \includegraphics[scale=.25]{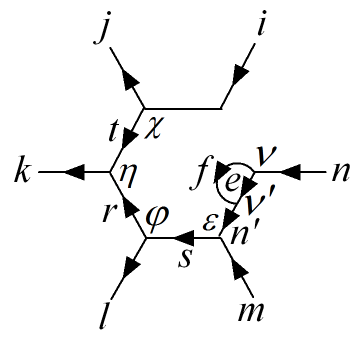} \epm\nonumber\\
&=& \sum_{t\chi\delta}\sum_{r\kappa\eta}\sum_{s\rho\varphi}\sum_{\nu^\prime\epsilon}(\cH^{jit,\chi\delta}_{fba,\al\bt})^\dagger \cF^{tfb,\delta\gamma}_{ckr,\kappa\eta}\t \cH^{rfc,\kappa\lambda}_{dls,\rho\varphi}
(\cF^{fen,\nu^\prime\epsilon}_{msd,\mu\rho})^\dagger \cO^{fe,\nu\nu^\prime}_n 
\Psi_\text{fix} \bpm \includegraphics[scale=.25]{Ha6} \epm \nonumber
\end{eqnarray}
Therefore, we finally derive:
\begin{align}
(\cU_p)^{a\al,b\bt,c\ga,d\la,e\mu,f\nu } _{trs,\chi\eta\varphi\epsilon} (i,j,k,l,m,n)=(\cH^{jit,\chi\delta}_{fba,\al\bt})^\dagger \cF^{tfb,\delta\gamma}_{ckr,\kappa\eta}\t \cH^{rfc,\kappa\lambda}_{dls,\rho\varphi}
(\cF^{fen,\nu^\prime\epsilon}_{msd,\mu\rho})^\dagger \cO^{fe,\nu\nu^\prime}_n 
\end{align}
\end{widetext}

Also $\cU_P$,
containing only one O-move (see Fig. \ref{hextotree}),
has a form $\cU_P=\cU_1\cP \cU_2$ where $\cU_{1,2}$ are
unitary matrices and $\cP$ is a projection matrix.  So the
rank of $\cB$ is equal or less than $D_{ijklmn}$.  Since it
is the identity in the $D_{ijklmn}$-dimensional space
$V_{ijklmn}$, the matrix $\cB$
is a hermitian projection matrix onto the space $V_{ijklmn}$:
\begin{widetext}
\begin{align}
&\ \ \
\sum_{a"\al"}
\sum_{b"\bt"}
\sum_{c"\ga"}
\sum_{d"\la"}
\sum_{e"\mu"}
\sum_{f"\nu"}
 \cB^{a\al,b\bt,c\ga,d\la,e\mu,f\nu } _{
a"\al",b"\bt",c"\ga",d"\la",e"\mu",f"\nu" } (i,j,k,l,m,n)
 \cB^{a"\al",b"\bt",c"\ga",d"\la",e"\mu",f"\nu" } _{
a'\al',b'\bt',c'\ga',d'\la',e'\mu',f'\nu' } (i,j,k,l,m,n)
\nonumber\\
&=
 \cB^{a\al,b\bt,c\ga,d\la,e\mu,f\nu } _{
a'\al',b'\bt',c'\ga',d'\la',e'\mu',f'\nu' } (i,j,k,l,m,n)
\end{align}
\end{widetext}

In the above calculation of the $\cB$, we first insert a
bubble on the $a$-link.  We may also calculate $\cB$ by first
inserting a bubble on other lines. All those different
calculations will lead to the same $\cB$ matrix, as
discussed above.

We note that $\hat B_{\v p_1}\hat B_{\v p_2}$ and $\hat B_{\v
p_2}\hat B_{\v p_1}$ are generated by different combinations
of F-moves and O-moves.  Since the two combinations
transform between the same pair of states, they give rise
to the same relation between the two states. Therefore $\hat
B_{\v p_1}$ and $\hat B_{\v p_2}$ commute
\begin{align}
 \hat B_{\v p_1}\hat B_{\v p_2} =\hat B_{\v p_2}\hat B_{\v p_1}
.
\end{align}
We see that the corresponding Hamiltonian $\hat H$ is a sum
of commuting projectors and is exactly soluble.

\bibliography{../../../bib/wencross,../../../bib/all,../../../bib/publst}

\end{document}